\documentclass[ApJ]{emulateapj}

\newcommand{\chandra}{{\it Chandra}}

\newcommand{\xmm}{{\it XMM}}
\newcommand{\suzaku}{{\it Suzaku}}
\newcommand{\fermi}{{\it Fermi}}

\newcommand{\red}{\textcolor{red}}

\newcommand{\gamray}{$\gamma$-ray}

\def\lsim{\raise0.3ex
  \hbox{$<$\kern-0.75em\raise-1.1ex\hbox{$\sim$}}\,}

\def\gsim{\raise0.3ex
  \hbox{$>$\kern-0.75em\raise-1.1ex\hbox{$\sim$}}\,}

\usepackage{ulem,amssymb} 
\usepackage{color} 

\slugcomment{Accepted for Publication in The Astrophysical Journal}

\shorttitle{Investigating the Structure of Vela X}
\shortauthors{Slane et al.}

\begin{document}

\title{
Investigating the Structure of Vela X
} 

\author{P.~Slane \altaffilmark{1},
I. Lovchinsky \altaffilmark{1,2},
C. Kolb \altaffilmark{3}
S.~L.~Snowden \altaffilmark{4},
T. Temim \altaffilmark{5}
J. Blondin \altaffilmark{3}
F. Bocchino \altaffilmark{6},
M. Miceli \altaffilmark{6},
R.~A.~Chevalier \altaffilmark{7},
J.~P.~Hughes \altaffilmark{8},
D.~J.~Patnaude \altaffilmark{1},
and T.~Gaetz \altaffilmark{1}
}

\altaffiltext{1}{
Harvard-Smithsonian Center for Astrophysics, 
60 Garden Street, Cambridge, MA 02138, USA;
slane@cfa.harvard.edu}

\altaffiltext{2}{
Department of Physics, Harvard University, USA
}

\altaffiltext{3}{
North Carolina State University, USA
}

\altaffiltext{4}{
Laboratory for High Energy Astrophysics, Code 662, NASA/GSFC, Greenbelt, MD,
20771, USA}

\altaffiltext{5}{
Space Telescope Science Institute, 3700 San Martin Drive, Baltimore, MD
21218, USA
}

\altaffiltext{6}{
INAF -- Osservatorio Astronomico di Palermo, 
Piazza del Parlamento I, 90134 Palermo, Italy
}

\altaffiltext{7}{
Department of Astronomy, University of Virginia, P.O. Box 400325,
Charlottesville, VA 22904-4325, USA}

\altaffiltext{8}{
Department of Physics and Astronomy, Rutgers University,
Piscataway, NJ 08854-8019, USA}

\begin{abstract}
Vela~X is the prototypical example of a pulsar wind nebula whose
morphology and detailed structure have been affected by the interaction
with the reverse shock of its host supernova remnant. The resulting
complex of filamentary structure and mixed-in ejecta embedded in a
nebula that is offset from the pulsar provides the best example we
have of this middle-age state that characterizes a significant
fraction of composite SNRs, and perhaps all of the large-diameter
PWNe seen as TeV sources. Here we report on an {\it XMM-Newton}
(hereafter {\it XMM}) Large Project study of Vela~X, supplemented
by additional \chandra\ observations. Through broad spectral modeling
as well as detailed spectral investigations of discrete emission
regions, we confirm previous studies that report evidence for ejecta
material within Vela~X, and show that equivalent width variations
of O VII and O VIII are consistent with temperature maps within the
PWN that show low-temperature regions where the projected SNR
emission appears to dominate emission from the ejecta. We identify
spectral variations in the nonthermal emission, with hard emission
being concentrated near the pulsar. We carry out investigations of
the Vela~X ``cocoon'' structure and, with hydrodynamical simulations,
show that its overall properties are consistent with structures
formed in the late-phase evolution of a composite SNR expanding
into a surrounding medium with a density gradient, with ejecta
material being swept beyond the pulsar and compressed into an
elongated structure in the direction opposite the high external
density.
\end{abstract}

\keywords{acceleration of particles, shock waves,
ISM: supernova remnants, ISM: individual objects (Vela X)}

\section{Introduction}

Located at a distance of only $\sim 290$~pc \citep{Dodson_etal03},
the Vela supernova remnant (SNR; Figure 1) houses a young pulsar
that powers the extended pulsar wind nebula (PWN) Vela X. The PWN
extends to the south of the pulsar, apparently the result of an
asymmetric reverse shock (RS) interaction associated with a large-scale
density anisotropy surrounding the SNR \citep[e.g.,][]{Blondin_etal01}.
Observations across the electromagnetic spectrum have been used to
characterize the emission properties and overall structure of the
PWN, as described below.

Radio observations of the Vela SNR \citep[G263.9$-$3.3;][]{Milne68}
show a large ($\sim 6^\circ$ diameter) shell-type remnant with a
central flat-spectrum PWN (Vela X). Studies of neutral hydrogen in
the Vela direction reveal a thin shell surrounding the SNR, with a
density $n_0 \approx 1 {\rm\ cm}^{-3}$ \citep{Dubner_etal98}, while
X-ray observations of the SNR with the {\it ROSAT} observatory
\citep{Aschenbach_etal95} show a limb-brightened shell of thermal
X-rays whose brightest emission is in the northeastern hemisphere,
toward the Galactic plane.  An overall asymmetry of Vela is evident,
with the emission in the south/southwest extending to larger radii
(Figure 1). Distinct bowshock-like structures are observed along,
or just outside, the SNR shell, with subsequent studies confirming
that these appear to be high-velocity ejecta fragments that have
exited the SNR \citep[e.g.,][]{Tsunemi_etal99}.  \citet{Bocchino_etal99}
show that the emission in the northeast (NE) region is characterized
by two distinct temperature components ($kT_1 \approx 0.1$~keV,
$kT_2 \approx 0.5$~keV), possibly indicating expansion into a
surrounding medium with dense clouds and a low-density inter-cloud
medium (ICM; $n_{\rm ICM} \approx 0.01 - 0.1 {\rm\ cm}^{-3}$).
Using \xmm\ observations, \citet{Miceli_etal05} carried out
observations of shock-cloud interactions along the northern rim of
Vela and concluded that the two distinct temperature components
reported in earlier studies are both associated with the clouds,
leading to an upper limit $n_{\rm ICM} < 0.06 {\rm\ cm}^{-3}$.
Optical and far-UV studies of Vela ``fragment D,'' located outside
the eastern limb, indicate interaction of this ejecta knot with an
external cloud with $n_{cl} \approx 4 - 11 {\rm\ cm}^{-3}$, providing
additional evidence for a higher overall density in the northern
and western regions of the SNR \citep{Sankrit_etal03}.

\begin{figure*}[t]
\includegraphics[width=7in]{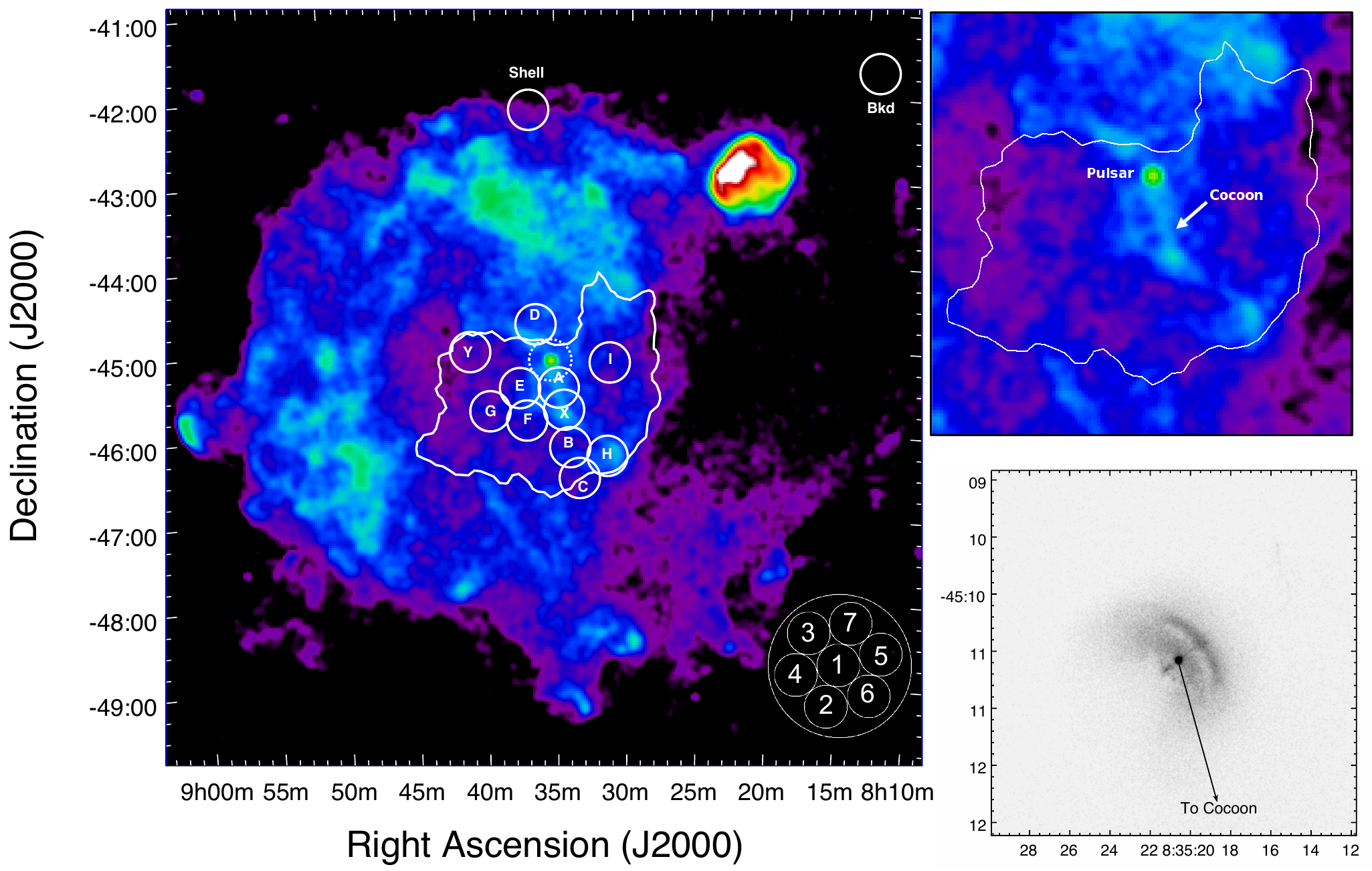}
\caption{
Left: {
\it ROSAT} image of the Vela SNR (G263.9$-$3.3). Soft emission is
shown in red and hard emission in blue/green. The contour is the
outer radio boundary of Vela X.  The individual circles identify
\xmm\ pointings discussed in the text. The dashed circular region
was taken in small-window mode and was used for imaging, but not
for spectral analysis. The numbered circles in 
the lower right illustrate regions used for spectral fitting of
each pointing (see Figure 6).
Upper right: Zoomed-in
region of {\it ROSAT} image showing pulsar and cocoon. The outermost
radio contour is shown to illustrate the extent of Vela~X.  Lower
right: \chandra\ image of the Vela pulsar and its surrounding compact
nebula.  The jet axis is in the SE-NW direction, roughly aligned
with the pulsar proper motion. The direction to the Vela~X cocoon
is indicated.
}
\label{fig1}
\end{figure*}

A $^{12}{\rm CO}$ survey of molecular clouds (MCs) in the vicinity
of the Vela SNR \citep{Moriguchi_etal01} establishes the presence
of a high MC concentration outside the northern limb of the remnant,
but little CO in the west/southwest where the X-ray emission is
more extended. This suggests an inhomogeneous density distribution in
the pre-explosion environment with an inter-cloud density $n_{\rm ICM}
\approx 0.01{\rm\ cm}^{-3}$, similar to that inferred from X-ray
studies, although the presence of an HI shell indicates a cooling shock
that requires a considerably higher density given the observed SNR
radius, illustrating that the medium surrounding the SNR is complex.

The Vela pulsar, located in the central regions of the SNR, has a
spin period of 89.3 ms, a characteristic age $\tau_c = 11.3$~kyr, and
a spin-down power $\dot{E} = 7 \times 10^{36} {\rm\ erg\ s}^{-1}$.
\chandra\ observations show that the pulsar is surrounded by a
compact nebula (Figure 1, lower right) with distinct features corresponding to
an inclined jet-torus structure \citep{Helfand_etal01}.  VLBI parallax
measurements \citep{Dodson_etal03} establish a
distance $d = 287^{+19}_{-17}$~pc and a proper motion that, when
combined with the torus inclination, establishes a pulsar velocity
of $\sim 80 {\rm\ km\ s}^{-1}$. The proper motion is along the
direction of the pulsar jet axis, providing evidence for alignment
of the kick velocity with the pulsar spin axis.


Radio observations of Vela~X reveal a morphology concentrated to
the south of the pulsar itself, suggesting that the SNR RS has
propagated more rapidly from the northern direction due to a higher
ambient density, thus leading to disruption of the northern part
of the PWN.  Higher resolution radio images \citep{Bock_etal98}
show a network of filamentary structure in the PWN, possibly formed
by Rayleigh-Taylor (R-T) instabilities in this interaction with the
RS (Figure 2).  ROSAT observations of the Vela~X region \citep{MO95}
reveal a large emission structure -- the so-called ``cocoon'' --
extending $\sim 45$~arcmin to the south of the pulsar (seen as a
distinct blue structure in Figure 1, upper right). The region is
characterized by a hard spectrum and appears to lie along a bright
elongated radio structure \citep{Frail_etal97}.  ASCA observations
established a two-component X-ray spectrum with the hard component
adequately described by either a power law or a hot thermal plasma
\citep{MO97}. Radio polarization maps show that the magnetic field
in Vela~X is roughly aligned with the cocoon in the central region
\citep{Bock_etal02}.

\begin{figure*}[t]
\centerline{\includegraphics[width=6.5in]{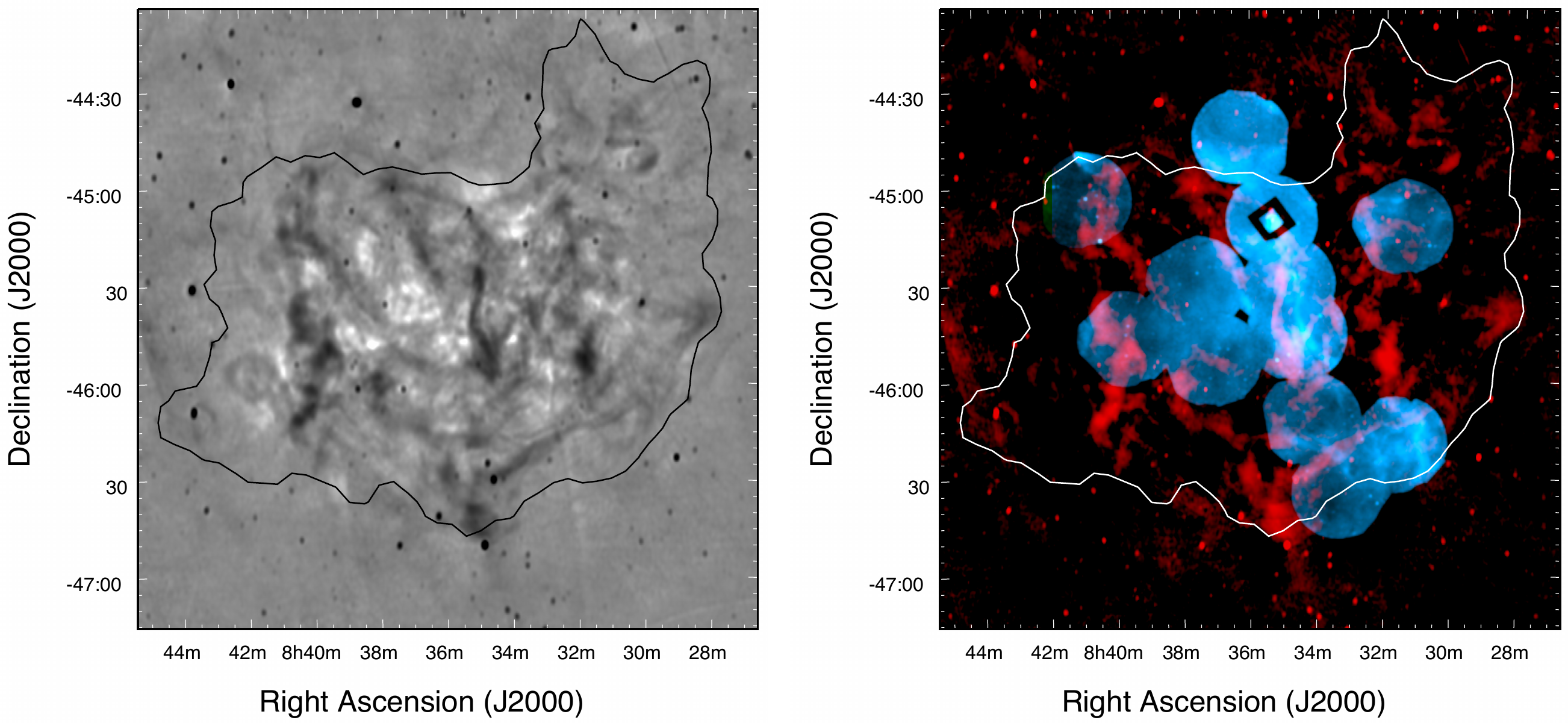}}
\caption{
Left: MOST radio image of Vela X showing filamentary structure
within the extended nebula. The single contour is the outermost
contour from the diffuse radio emission in Vela X. White regions in the image 
correspond to negative fluxes associated with missing coverage
due to a lack of small baselines. Right: Exposure-corrected mosaic of
\xmm\ pointing in Vela~X, shown in blue, along with the MOST radio
image, shown in red.  The bright, elongated radio structure in the
central region lies adjacent to the X-ray cocoon.
}
\label{fig2}
\end{figure*}

Using the {\it CANGAROO} telescope, \citet{Yoshikoshi_etal97}
detected $\gamma$-ray emission at energies above 2.5~TeV, located
somewhat southeast of the pulsar. Assuming inverse-Compton scattering
off of the same electron population responsible for the hard X-ray
spectrum, they estimated a magnetic field strength $B \lsim 4 \mu$G.
Subsequent observations with {\it H.E.S.S.} identified a TeV nebula
larger than the X-ray cocoon, with a spectrum consistent with either
a broken power law or a single power law with an exponential cutoff
\citep{Aharonian_etal06}. The brightest region of TeV emission is
concentrated along the X-ray cocoon. Using \xmm\ observations along
the central region of the cocoon, \cite{LaMassa_etal08} detected
two distinct emission components -- a power law with a spectral
index of $\sim 2.2$ and a thermal plasma with enhanced abundances
of O, Ne, and Mg, presumably associated with ejecta that have been
mixed into the PWN upon interaction with the RS. Broadband spectral
modeling showed that the radio, nonthermal X-ray, and TeV $\gamma$-ray
emission can be understood as the result of synchrotron and IC
emission, but with a broken power law or a distinct population of
radio-emitting electrons. A magnetic field strength of $\sim 5
{\mu}$G was estimated, consistent with the value estimated from TeV
studies.  Modeling by \cite{deJager_etal08} predicted observable
GeV emission from Vela X based on a leptonic model for the TeV
emission.  The PWN is observed at energies up to $\sim 200$~keV
with {\it BeppoSAX} \citep{Mangano_etal05}, and observations with
\fermi\ reveal \gamray\ emission extended over the entire radio-emitting
region, but with a centroid that is distinctly offset toward the
west from the peak of the TeV emission \citep{Abdo_etal10}.  X-ray
observations with \suzaku\ \citep{Katsuda_etal11} identify nonthermal
emission in the northeast that appears to extend beyond the radio
emission, leading to the suggestion that perhaps the X-ray and
radio-emitting components are associated with different electron
populations.

Here we describe X-ray observations carried out in an \xmm\ Large
Project to study Vela X, along with \chandra\ observations coincident
with the early region studied by LaMassa et al. (2008). We apply
a hydrodynamical model, constrained by the observed properties of
Vela~X, to investigate the formation of the cocoon structure.  In
Section 2, we describe our observations and data reduction. The
results of our spectral analysis are presented in Section 3, and
our hydrodynamical simulations of the Vela SNR and Vela~X are
described in Section 4.  We present a discussion of our results in
the context of the multiwavelength data on Vela~X in Section 5, and
summarize our conclusions in Section 6.

\section{Observations and Data Reduction}

\subsection{\xmm}
Vela~X was observed for a total of 371~ks in eight pointings as an
\xmm\ Large Project carried out between 30 April and 21 Dec 2009.
The individual pointings are identified as regions A-H in Figure
1, and the exposure times are summarized in Table 1. An additional
55 ks observation (region I) was obtained on 22 May 2011 at a
position corresponding to the peak of the GeV emission measured
with the \fermi-LAT \citep{Abdo_etal10}. The observations were carried
out in full frame mode.

\begin{table}[t]
\begin{center}
\caption{\footnotesize{Vela~X Observations}}
\label{tab:obs}
\begin{tabular}{ccccc}
\toprule
\noalign{\smallskip}
\noalign{\smallskip}

		&		&Exp.		&Clean Exp. \\
ObsID           &Pointing        &(ks)		&(M1/M2/pn)	&Date \\
\\
\multicolumn{5}{c}{\sl XMM-Newton} \\
0603510101      &A               &25		&15.6/15.8/8.3		&2009-06-30 \\
0603510201      &B               &67		&51.5/52.5/37.7		&2009-04-30 \\
0603510301      &C               &90		&87.3/89.9/62.7		&2009-12-21 \\
0603510401      &D               &20		&22.6/22.8/18.5		&2009-10-17 \\
0603510501      &E               &38		&39.6/39.7/32.8		&2009-05-20 \\
0603510601      &F               &41		&43.3/43.7/31.5		&2009-05-10 \\
0603510701      &G               &68		&55.5/63.9/34.0		&2009-06-09 \\
0603510901      &H               &41		&42.2/4.3.3/28.9	&	2009-06-15  \\
\\
0672040101      &I               &55		&53.1/53.9/31.2		&2011-05-20 \\
\\
0094630101      &X               &21		&20.6/20.4/16.2		&2002-05-02 \\
0506490101      &Y               &45		&34.0/30.4/21.3		&2007-05-29 \\
0153350101      &bkdg            &13.9		&12.6/12.0/9.5		&2002-10-14 \\
0203960101      &shell           &27		&26.6/26.9/15.1		&2004-10-30 \\
\\
\multicolumn{5}{c}\chandra \\
12697		&X		&25		&23.1			& 2010-09-28\\

\noalign{\smallskip}
\hline
\noalign{\smallskip}
\end{tabular}
\end{center}
\end{table}

\begin{figure}[t]
\epsscale{1.00}
\centerline{\includegraphics[width=3.0in]{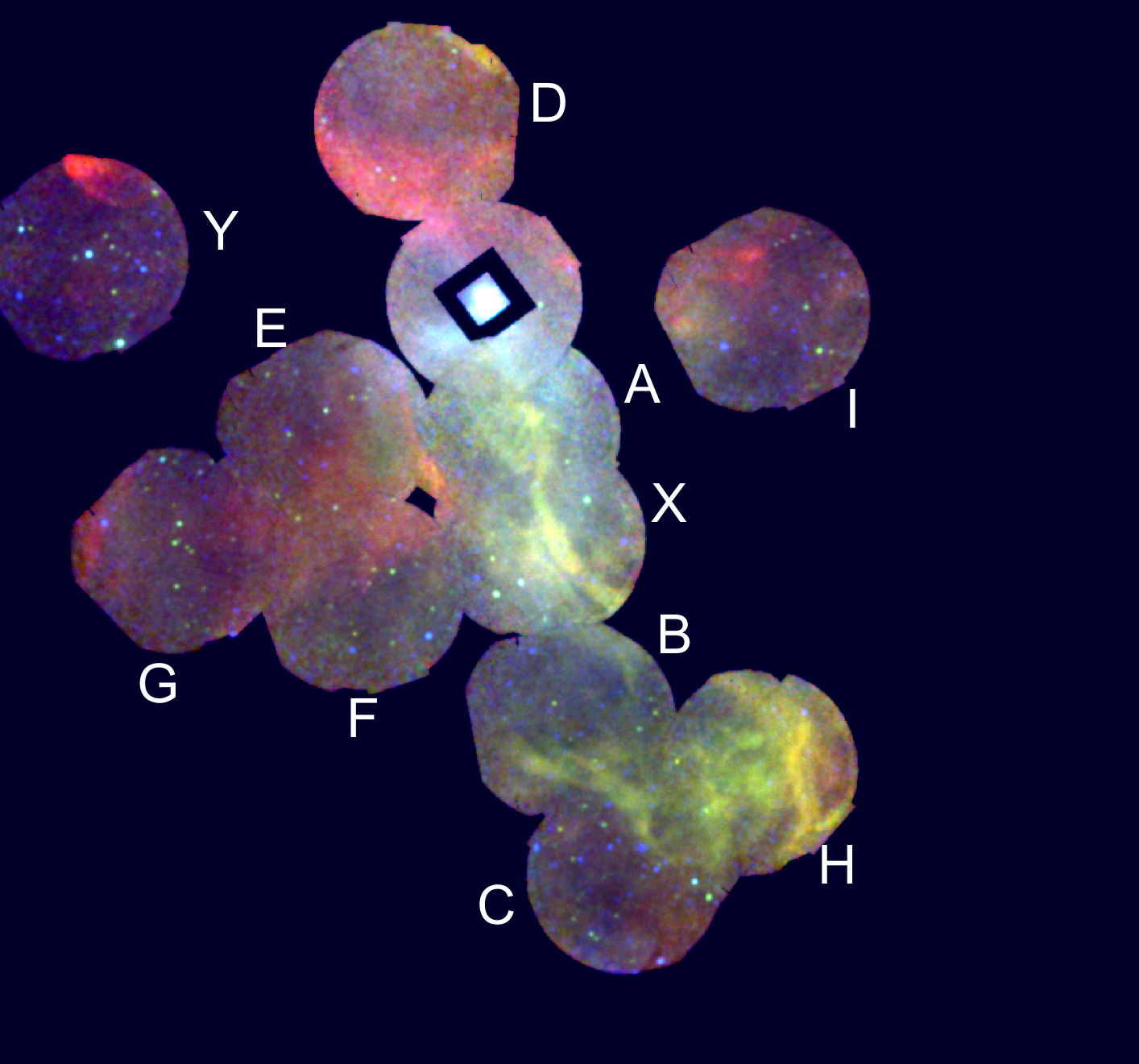}}
\caption{
Exposure-corrected mosaic of Vela~X regions covered by \xmm, with
red/green/blue representing energy ranges 0.4-0.75/0.75-1.4/2.0-7.2~keV.
The cocoon structure is evident as region of harder emission extending
to the south/southwest from the pulsar, located within the small
window in an upper pointing. (Note that the pointing containing the
pulsar was not used in subsequent spectral analysis because of
incomplete coverage due to the small window mode.)
}
\label{fig3}
\end{figure}

Additional archival observations from the interior of Vela X (regions
X and Y) were also included in the analysis. A pointing at the
pulsar, taken in Large Window mode (dotted circle in Figure 1) was
included in images of the system, but was not used in the spectral
analysis.  A pointing along the northern shell of the Vela SNR
(``Shell'' region) was used to constrain the column density and
thermal contributions from the SNR rim. The SNR spectrum was extracted
from a circular region with a 4.3~arcmin radius at the center of
the detector, while a pointing outside the SNR was used to obtain
a background spectrum (``Bkd'' pointing). All data were reprocessed
and cleaned with {\it SAS Version 15.0.0}. The cleaned exposure
times for each EPIC detector are shown in Table 1.

An exposure-corrected image from the \xmm\ observations is shown
in Figure 2 (right), overlaid on the radio image. The extended X-ray
cocoon resides along, but distinct from, an extended radio structure.
A three-color image from the \xmm\ mosaic is shown in Figure 3. The
cocoon structure is evident as a long feature that extends toward
the southern regions of the PWN. Distinct filamentary structures
are also observed in the south, and there are noticeable spatial
variations in the soft emission identified in red. We discuss the
spectra from these regions in Section 3.

\subsection{\chandra}

A portion of Vela~X was observed with \chandra\ on 28 September
2010. The observation (ObsID 12697) was carried out with ACIS-I in
Very Faint mode, with the pointing centered on the position of \xmm\
field X (see Figure 1). Data were reprocessed and cleaned with {\it
Ciao Version 4.8} resulting in a total exposure of 23.1~ks.

\begin{figure}[t]
\epsscale{1.00}
\includegraphics[width=3.4in]{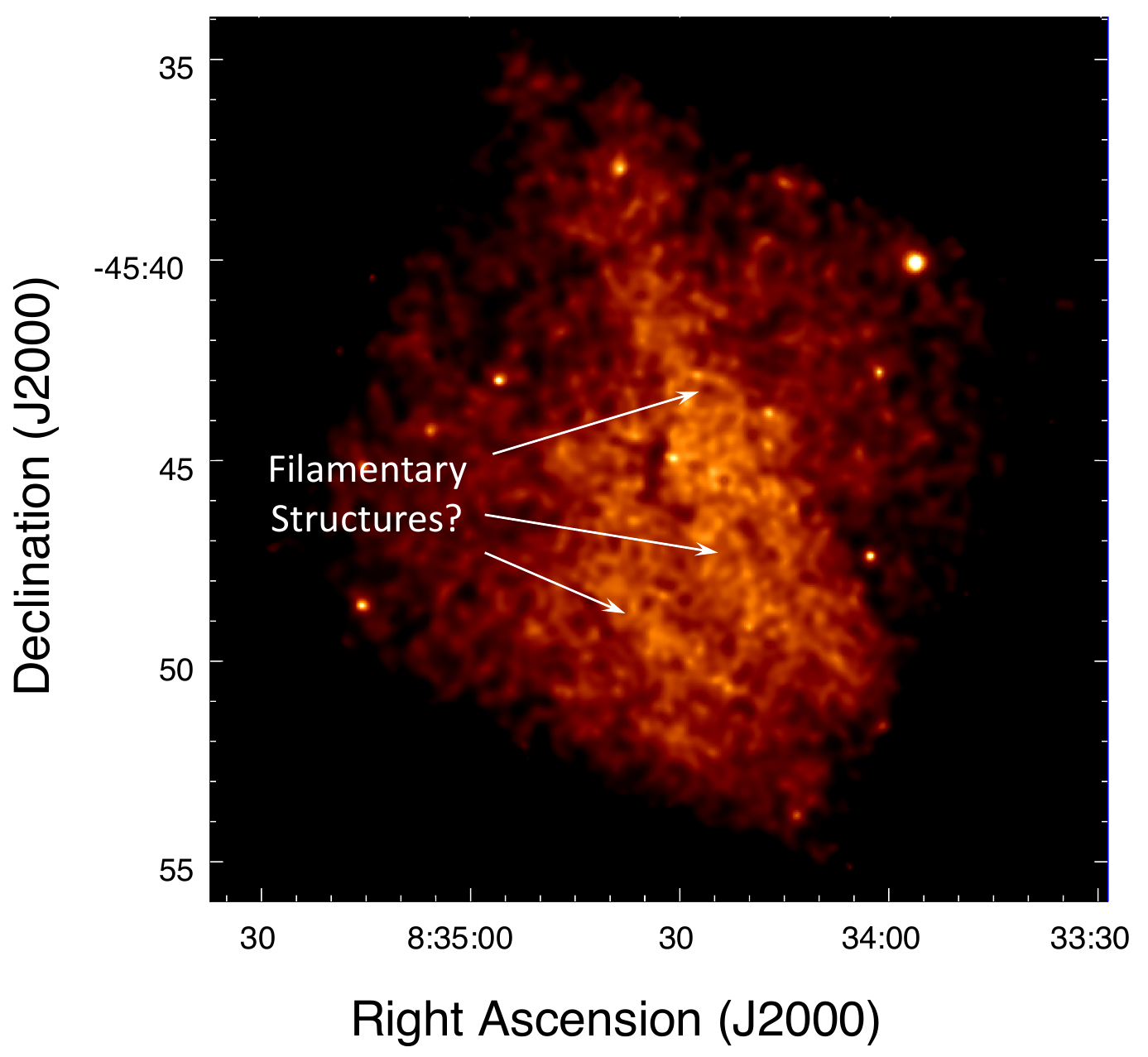}
\caption{
Exposure-corrected, adaptively-smoothed ACIS image of the central
portion of the Vela~X cocoon (largely overlapping Region X in Figure 1). 
The structure is primarily composed
of confined diffuse emission, although there is some evidence of
narrow filamentary structures within the cocoon.
}
\label{fig4}
\end{figure}

In order to investigate the spatial structure from this region of
the Vela~X cocoon, we created an adaptively-smoothed, exposure-corrected
image using a Gaussian smoothing kernel with a minimum of 100 counts
on a logarithmic size scale ranging from 0.1 to 50 pixels (using a
bin size of 4 ACIS pixels, or approximately $2^{\prime\prime} \times
2^{\prime\prime}$). The image is shown in Figure 4. The overall
extended morphology is similar to that observed with \xmm, although
there is some evidence for filamentary structure in the X-ray emission.

\begin{figure*}[t]
\centering
\setlength\fboxsep{0pt}
\setlength\fboxrule{0.0pt}
\fbox{
\includegraphics[height=3in]{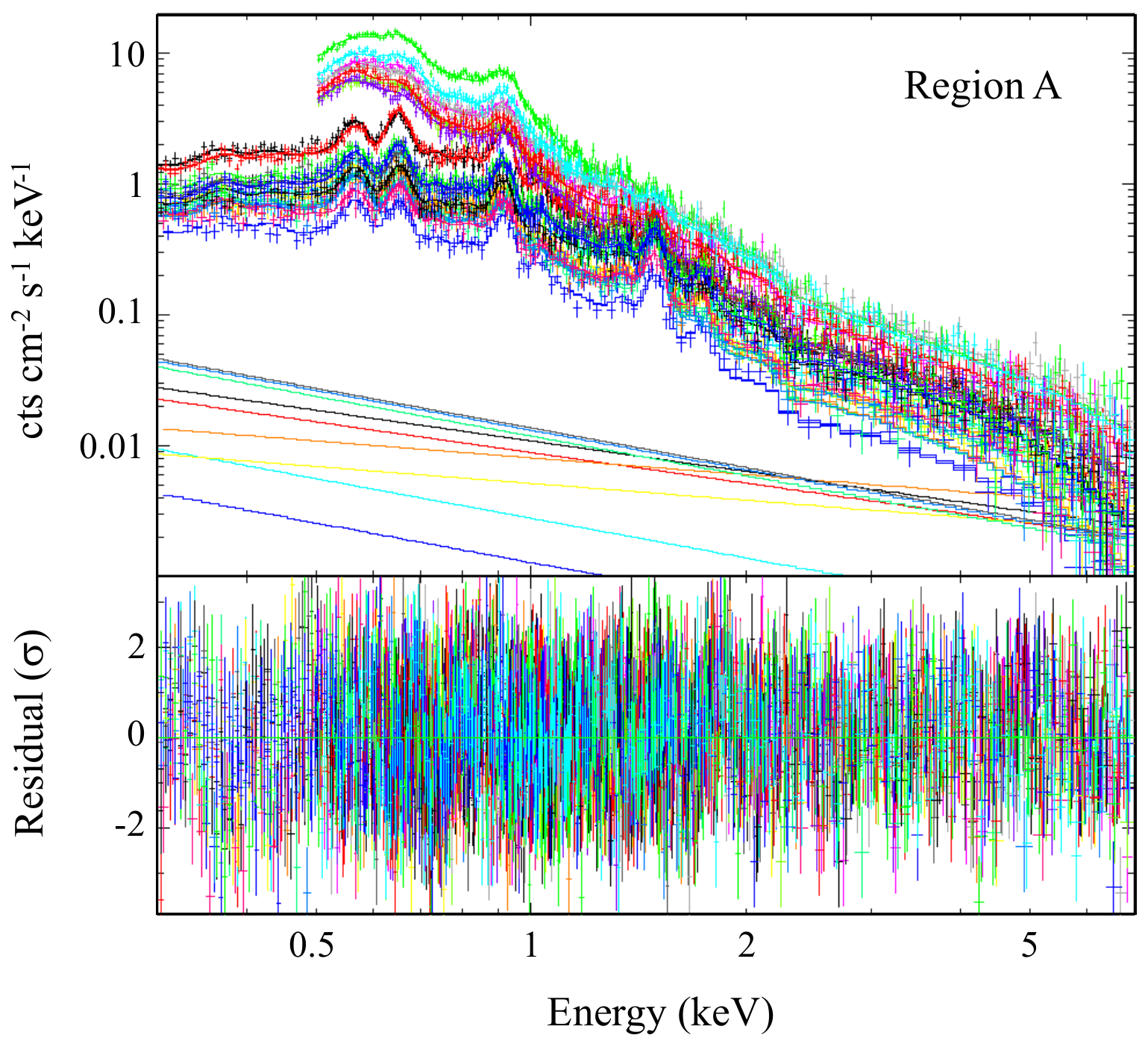}
\includegraphics[height=3in]{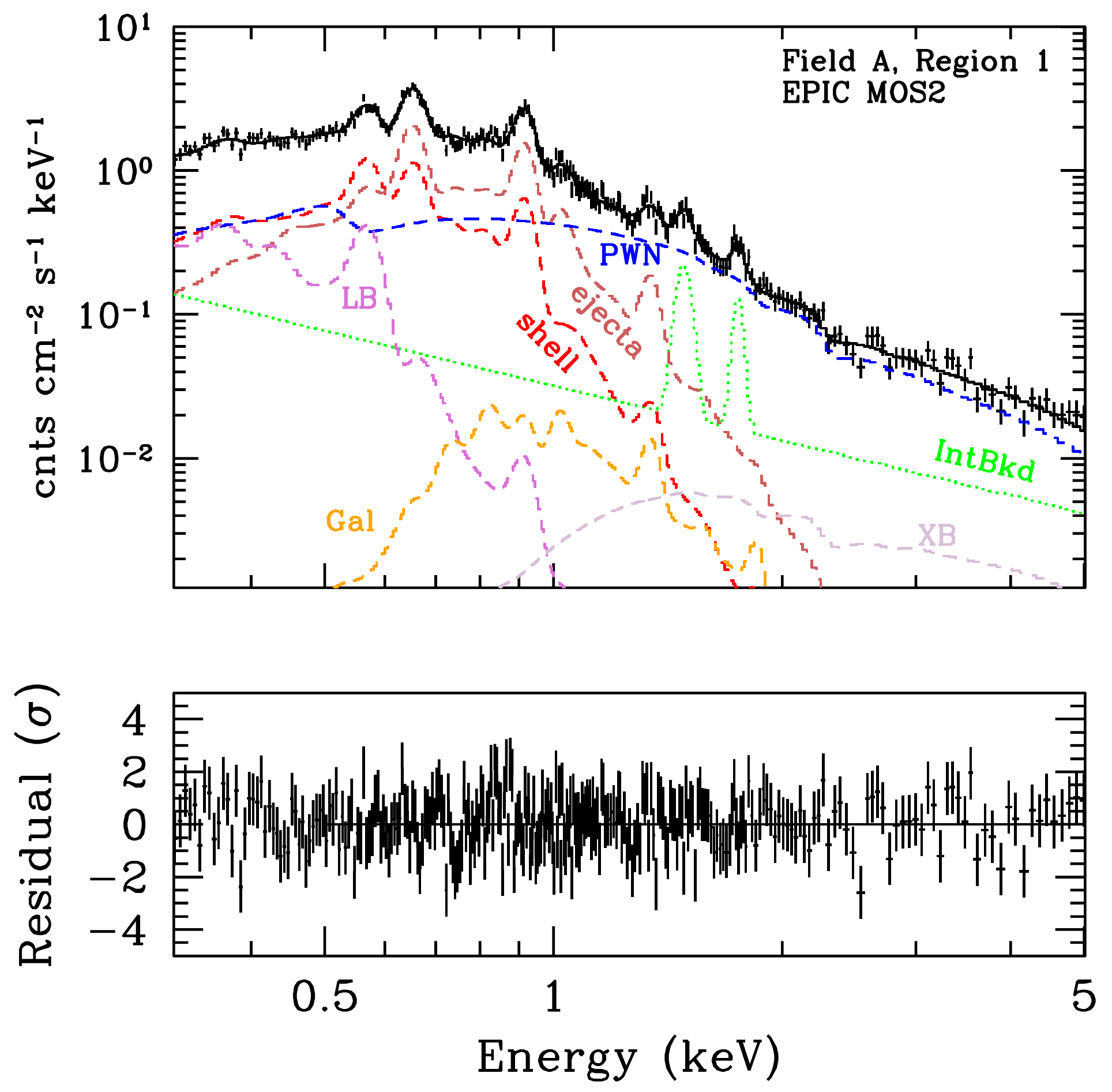}
}
\caption{
Left: Joint fit to all MOS and pn spectra from Region A. The reduced
chi-squared for the joint fit is $\chi^2_r = 1.2$. The lines in the lower
portion of the plot correspond to the power law component of the
internal background model.
Right: MOS2 spectrum of center region from Field A showing model components
described in text. 
}
\label{fig5}
\end{figure*}

\section{Spectral Analysis}

To investigate the X-ray emission from Vela~X, we concentrated on
searches for nonthermal emission from the PWN along with thermal
emission from any ejecta component that may have been mixed into
the nebula by the reverse shock interaction. (Projected emission
from the SNR shell was modeled as well, as discussed below.)
Spectral modeling
was carried out using {\sc xspec 12.9.0d} with {\sc atomdb 3.03}.
For each of the eleven
\xmm\ pointings described in Table 1 (excluding the background and
shell pointings), we defined seven circular
regions (in detector coordinates), with radii of 4.3 arcmin, that 
approximately covered the
detector field.  Due to the large size of Vela~X, and of the Vela
SNR, local background subtraction is not possible. We thus followed
the treatment for extended source analysis described in the \xmm\
ESAS document.\footnote{\url{https://heasarc.gsfc.nasa.gov/docs/xmm/xmmhp\_xmmesas.html};
see also \citep{Snowden_etal08} and \citep{KS08}.} For each region
in each pointing, we eliminated point sources and then subtracted
an internal background contribution for each detector using filter
wheel closed data. We then defined a sky background model that
consists of contributions from the diffuse extragalactic background
(a power law with $\Gamma = 1.46$ and surface brightness $10.5 {\rm\
photons\ cm}^{-2} {\rm\ s}^{-1} {\rm\ keV}^{-1} {\rm\ sr}^{-1}$ at
1~keV); a nonequilibrium ionization (NEI) plasma model ({\sc vnei}
in {\sc xspec}) for the projected contribution from the Vela SNR
shell; 
an equilibrium plasma model ({\sc apec}) for the integrated
contributions from hot gas in the Galactic plane; and an unabsorbed
equilibrium plasma model for emission from the Local Bubble.
Absorption was modeled with the {\sc tbabs} model, and abundances
were set to those from \citep{Wilms_etal2000}. Residual internal
background was modeled with a power law with additional Gaussian
lines and 1.48 keV and (for MOS only) 1.75 keV corresponding to
fluorescence from Al and Si; these components were not folded through
the effective area (arf) file. For each region, the free parameters
for the non-internal background components were fixed to be the
same for each detector (but allowed to vary from region to region).
The contributions from Vela~X were modeled by an absorbed power law
and an NEI plasma with variable abundances, corresponding to emission
from the PWN and from ejecta. The absorption for the Vela shell,
PWN, and ejecta components were tied together, as was the absorption
for the Galactic thermal emission and extragalactic X-ray background
components.

To establish the properties for the SNR shell, we first modeled the
spectrum from the Shell region identified in Figure 1, using the
above model without the Vela~X contributions. We simultaneously fit
the spectrum from the Bkd region, using only the background components
described above. We found that the remnant shell is adequately
described ($\chi^2_r = 1.2$) by an NEI model with $kT_{SNR} =
0.22$~keV with a neon abundance [Ne] = 2.8 relative to ISM values,
an ionization timescale $\tau = n_e t \sim 10^{13} {\rm\ cm^{-3}
s}$ (i.e., effectively in equilibrium, possibly suggesting a high
density component that could be associated with clumps, although $\tau$
is not well-determined around equilibrium, and could as low as
$\sim 2 \times 10^{12} {\rm\ cm^{-3} }$) and an absorption column
$N_H = 10^{20}{\rm\ cm}^{-2}$, in good agreement with results from
\citet{Miceli_etal05}.  We fixed these values in our fits for the
Vela~X regions, along with the temperature of the Local Bubble
contribution ($kT = 0.1$~keV). The normalization of the projected
SNR emission within Vela~X was allowed to vary from region to region
to account for potential variations in the surface brightness.

\begin{figure*}[t]
\centering
\setlength\fboxsep{0pt}
\setlength\fboxrule{0.0pt}
\fbox{
\includegraphics[height=2.5in]{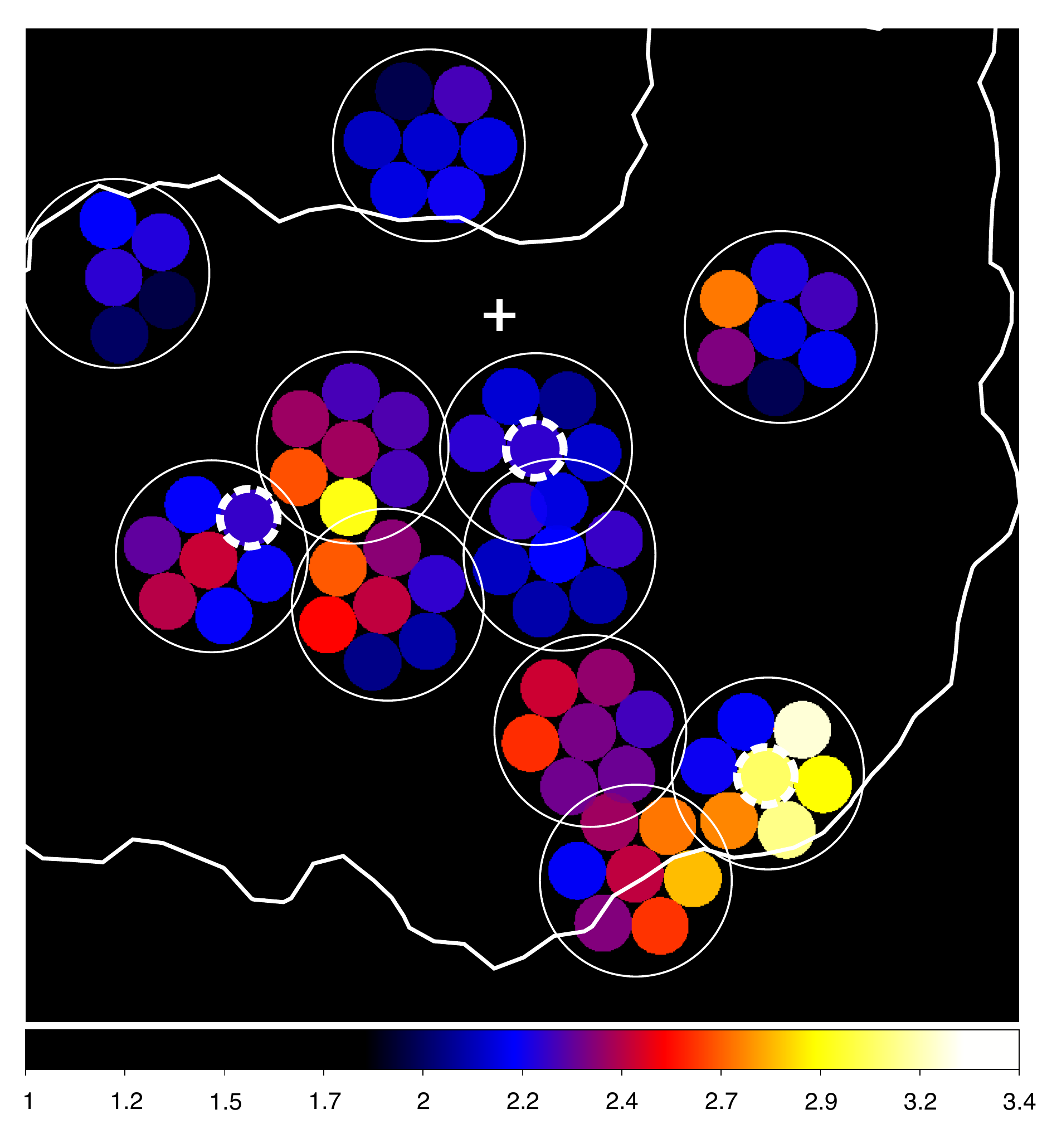}
\includegraphics[height=2.5in]{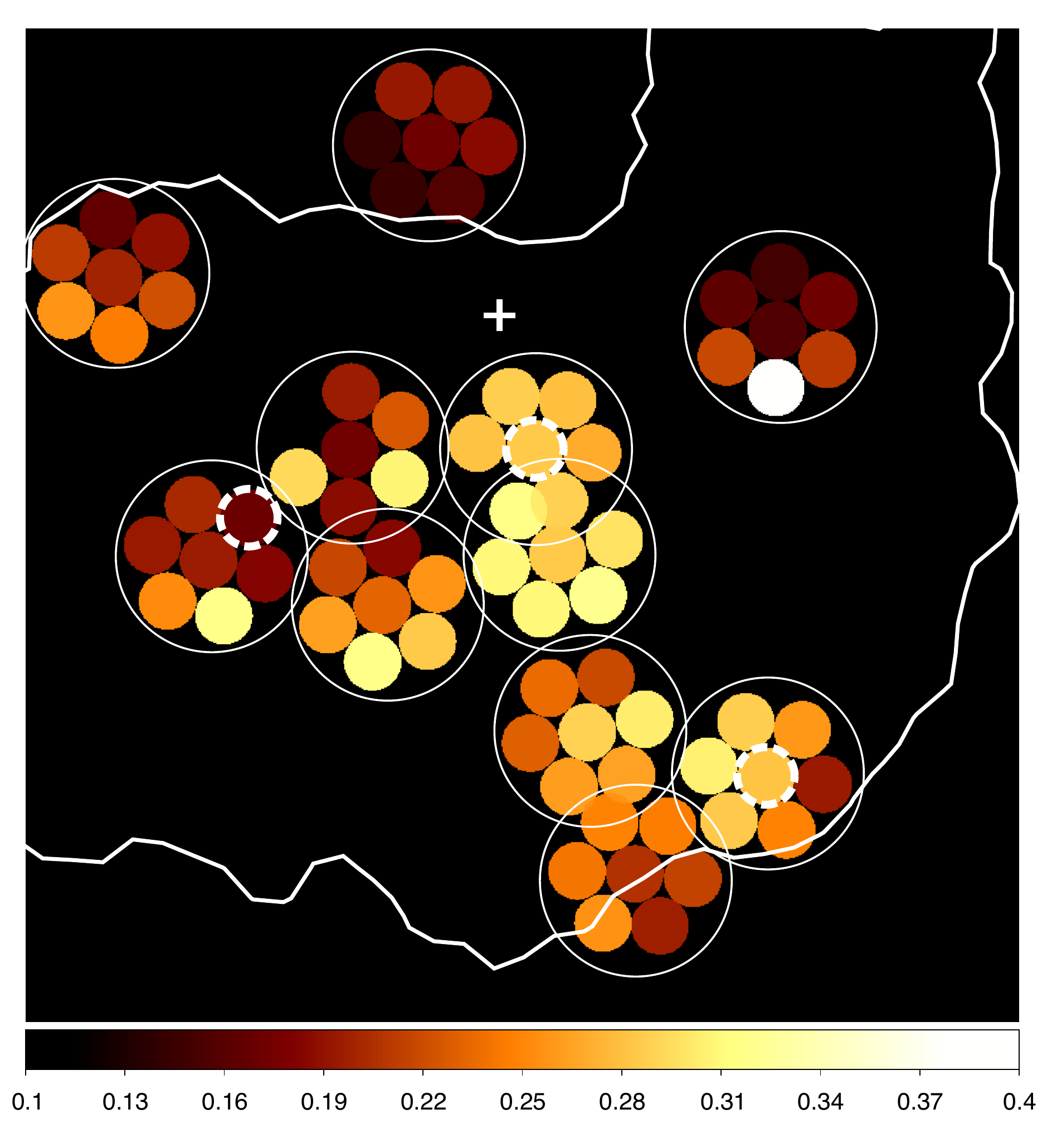}
\includegraphics[height=2.5in]{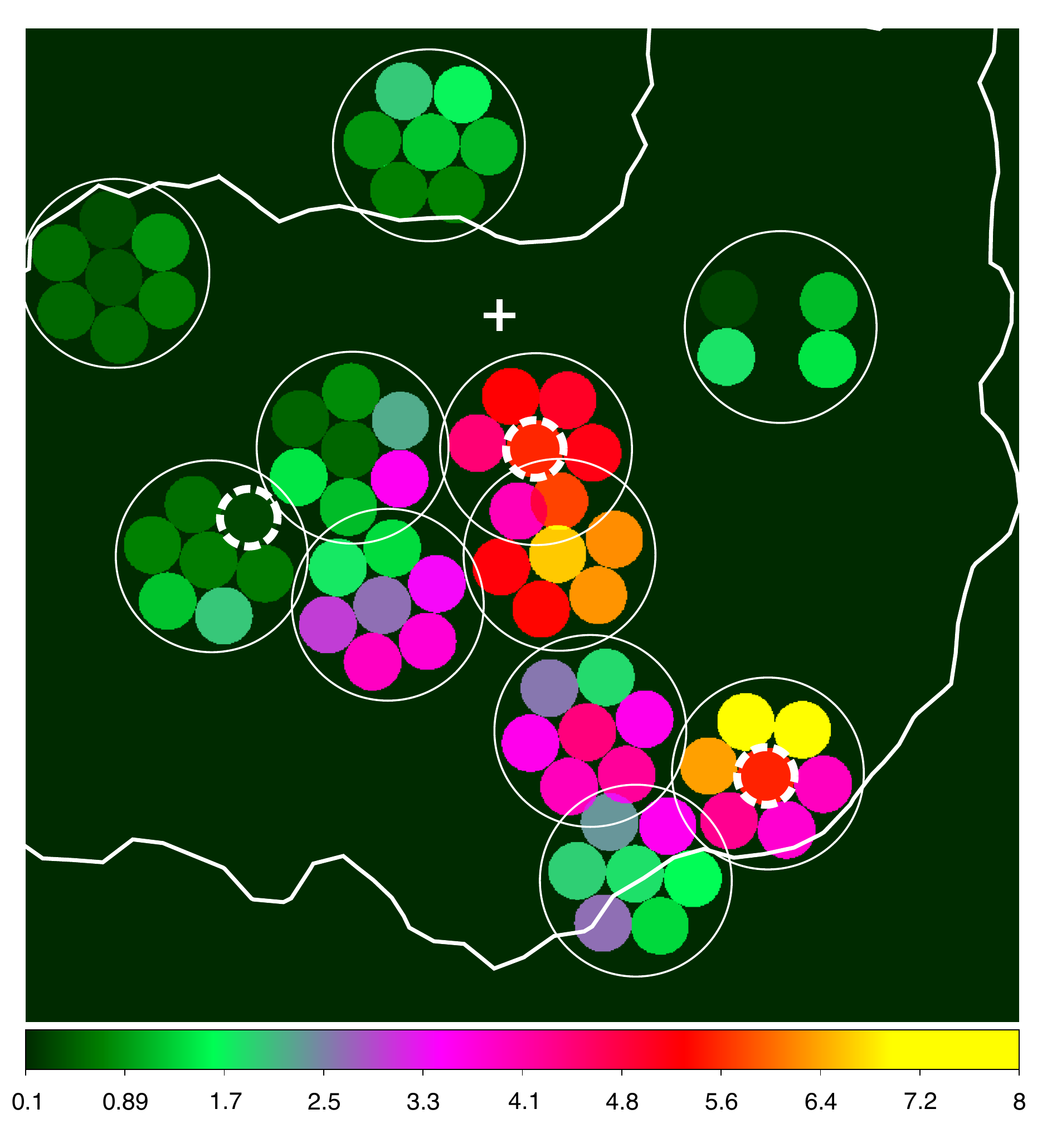}
}
\caption{
Maps of the power law index (left), temperature in keV
(middle), and Ne abundance relative to the ISM value (right)
from spectral fits of subregions in \xmm\ pointings in Vela~X. The
temperature and abundance correspond to the ejecta component in the
spectral model. The outermost radio contours provide the rough
outline of the PWN and the cross indicates the position of the
pulsar. The color bars indicate the values for the photon index,
the temperature in keV, and the abundance relative to solar values,
respectively. The parameter uncertainties vary from region to region,
but are typically $\lsim 0.05 - 0.1$ for the spectral index, $\sim
0.01$~keV for the temperature,  and $\sim 0.1 - 0.3$ for the
abundance. Distinct spectral steepening is observed along the cocoon,
which also shows higher temperatures and abundances than other
regions within the PWN. For reference, region 7 for each of the
pointings (see inset to Figure 1, left) is identified with a red
dot in the center panel. Dashed circles correspond to regions for
which spectra are shown in Figure 7.
}
\label{fig6}
\end{figure*}

For each \xmm\ pointing on Vela~X, we followed the same procedure,
treating each region independently, and simultaneously fitting the
background. The parameters for the sky background components and
for the column density for the Vela SNR were fixed at values
determined from the fit for the Shell region, except for the
normalizations (other than that for the diffuse extragalactic
component). In addition to the associated normalizations, the power
law index for the PWN component and the temperature and ionization
timescale for the ejecta component, along with its abundances of O
and Ne, were treated as free parameters.

The left panel of Figure 5 shows the joint fit for all regions in
Vela X Pointing A. While the overall fit is reasonably good ($\chi^2_r
= 1.2$), it is important to note that by fitting all spectra from
these regions simultaneously, we are potentially hiding some
variations in the individual spectra.  Our purpose here is to
establish the presence of emission from the PWN and ejecta, and to
search for global variations in the power law index, temperature,
and abundances. As discussed further below, the temperature of the
ejecta component is only a small amount higher than that of the SNR
shell, leading to a degeneracy between these components in regions
where the temperatures are very close, or where the ejecta component
is faint. The quality of the joint fits for the different regions
range between $\chi^2_r \approx 1.2 - 1.3$ along the cocoon, and
$\chi^2_r \approx 1.3 - 1.5$ in the outer regions. 

Figure 5 (right) illustrates the different model components for the
MOS2 spectrum from the central portion (Region 1) of pointing A.
The Galactic thermal (Gal) and extragalactic (XB) background
components contribute little to the overall emission, and the
emission from the Local Bubble contributes only at the lowest
energies, at and below the O VII line at 0.56~keV. The power law
component from the residual internal background is only significant
at high energies, but the two fluorescence lines are readily seen
in the full spectrum. For this region, nonthermal emission from the
PWN dominates above 1~keV, and the thermal emission from the ejecta
component shows significantly enhanced Ne, O, and Mg abundances.
The model fit yields $\chi^2_r = 1.0.$

\begin{figure*}[t]
\centering
\setlength\fboxsep{0pt}
\setlength\fboxrule{0.0pt}
\fbox{
\includegraphics[height=3in]{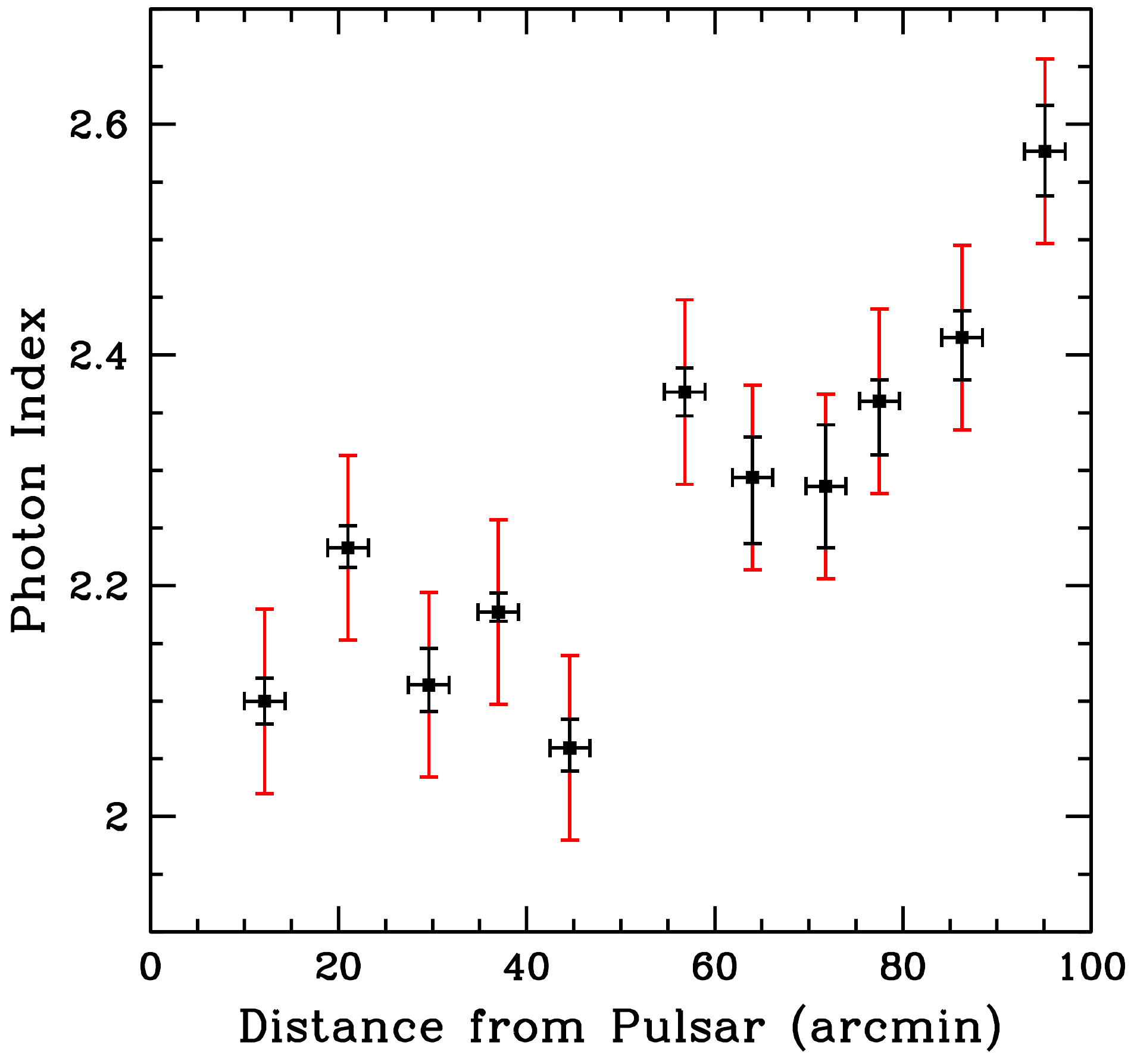}
\includegraphics[height=3.1in]{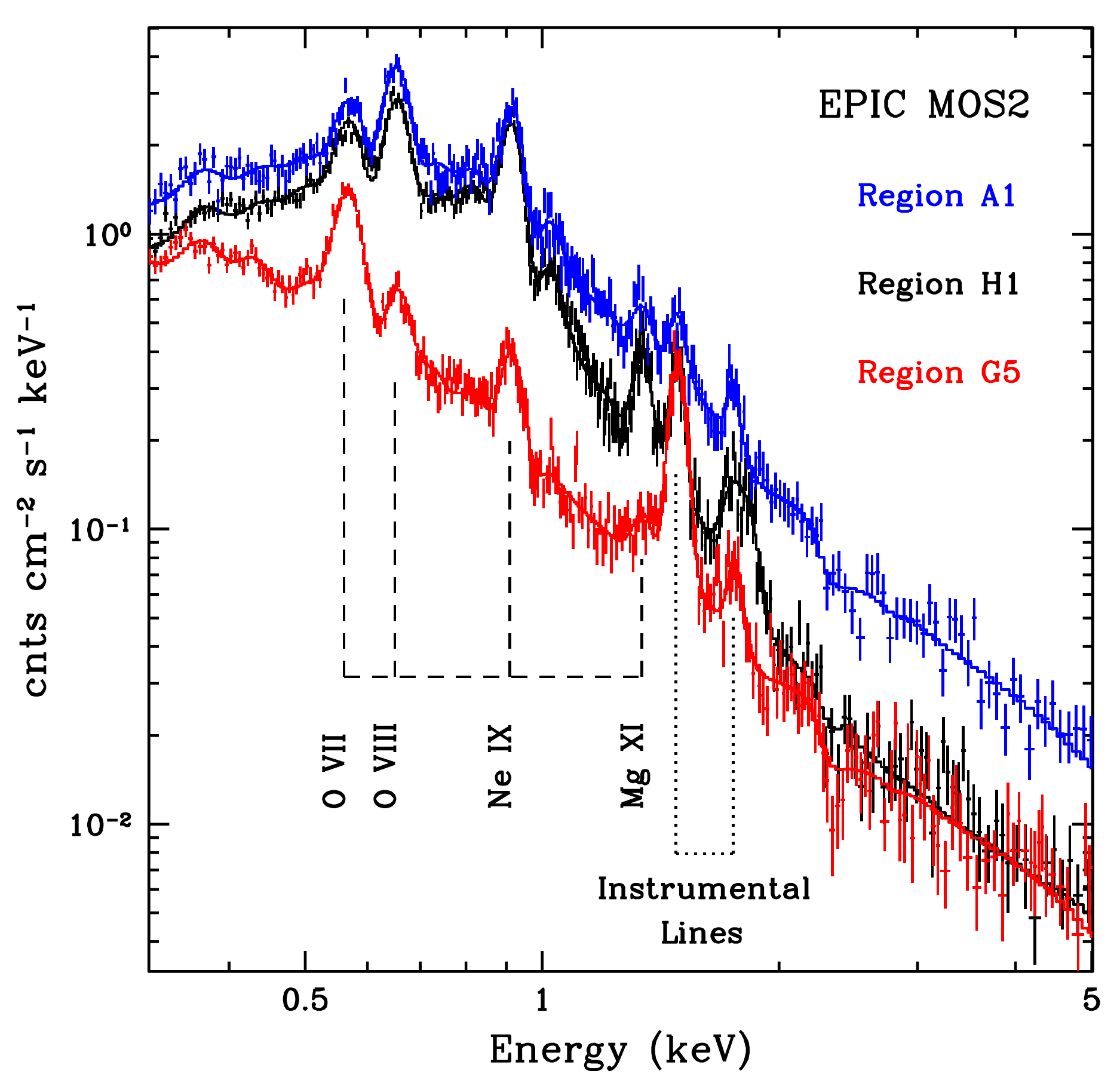}
}
\caption{
Left: Variation in spectral index with distance from pulsar along
Vela X cocoon. A distinct spectral steepening is observed at large
distances along the cocoon. Black error bars correspond to 90\% 
confidence uncertainties
from the joint fits while the red error bars represent systematic
errors estimated by comparing values from the fits with those
obtained by fitting individual regions from a given pointing
separately.
Right: Spectra from regions A1, H1, and G5 which show increasingly
lower temperatures. The reduced amount of O VIII (relative to O
VII) and Mg X1 emission in Region G5 provides particular evidence
for the lower temperature. (See also Figure 8.)
}
\label{fig7}
\end{figure*}

\begin{figure*}[t]
\centering
\setlength\fboxsep{0pt}
\setlength\fboxrule{0.0pt}
\fbox{
\includegraphics[width=3in]{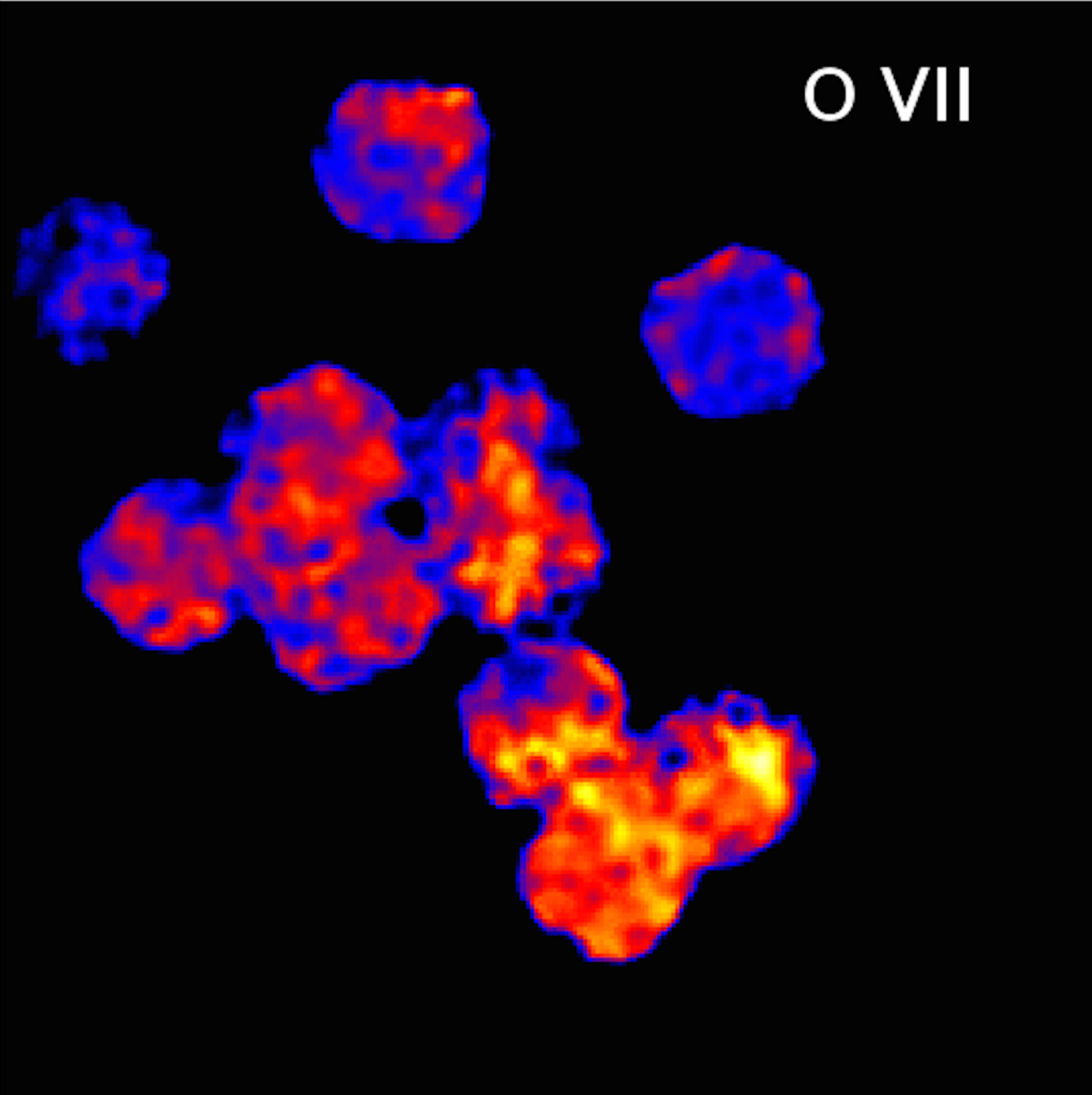}
\includegraphics[width=3in]{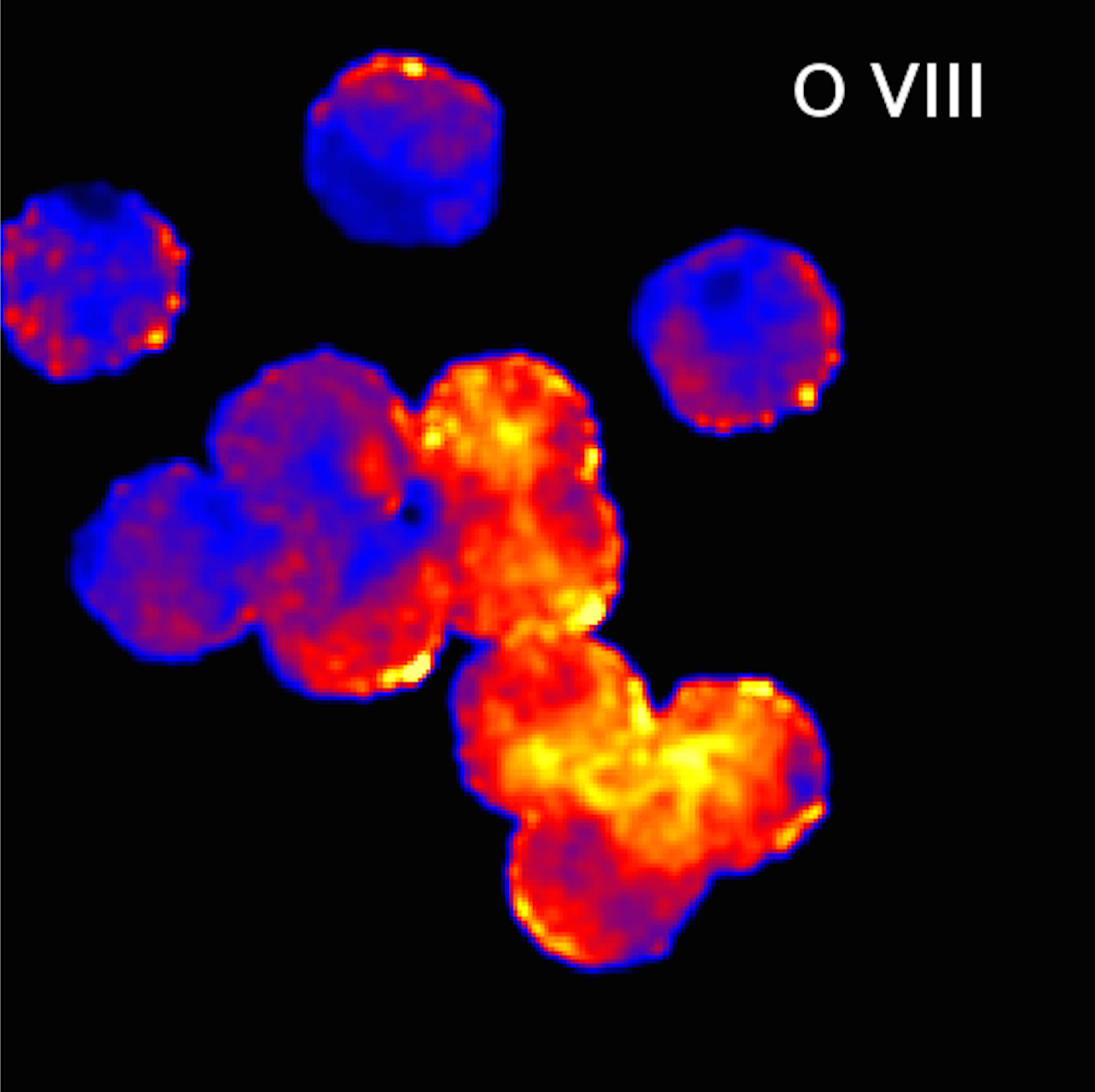}
}
\caption{
Equivalent-width maps (see text for definition) for O VII and O VIII
lines in Vela~X. Note the close correlation between the O VIII map
and the temperature map in Figure 6 (center). Edge brightening along
the outskirts of some regions is an artifact associated with the rapidly
declining effective area at the edge of the detector.
}
\label{fig8}
\end{figure*}

Nonthermal emission is detected throughout Vela~X, with the observed
spectral index varying spatially over a range $\Gamma \sim 1.9 -
3.2$, with the exception of Region Y which shows anomalously low
values in some regions (possibly related to a considerably higher
hard internal background component than observed in the other fields,
which may be compromising our ability to constrain any PWN emission).  
A spectral index map is shown in Figure 6
(left).  The spectrum is hardest near the pulsar, with a distinct
steepening at larger distances; regions adjacent to the pulsar show
$\Gamma \sim 2$ while those at the outer edges of the PWN show
values of $\Gamma \sim 3$. The spectrum is harder along the cocoon
region (and along the extended direction to the south) than in other
regions of Vela~X.  Interestingly, the spectrum in Region I,
corresponding to the peak of the \fermi-LAT emission, appears harder
than regions at similar distance on the other side of the cocoon.
The spectral index from regions along the direction of the cocoon,
and extending to the edge of the PWN in the south, along the
cocoon, is shown
in Figure 7 (left).  A significant steepening is evident at larger
distances, indicating that synchrotron losses are significant for
regions far from the pulsar. Uncertainties in the index vary from
region to region, but are typically $\lsim 0.05 - 0.1$ (90\% CL).

We detect a significant thermal component within Vela~X that appears
to be associated with hot ejecta, with enhanced abundances of Ne
and O in regions along the cocoon (as well as enhanced Mg in some
fits of individual regions).  The temperature varies throughout the
PWN ($kT \sim 0.14 - 0.32$) but is hottest along the filamentary
regions seen in Figure 3. A map of the temperature distribution is
shown in Figure 6 (center). The ionization timescale of this thermal
component is high throughout the PWN, consistent with the plasma
being in ionization equilibrium. The O and Ne abundances are highest
near the cocoon, suggesting that the ejecta component is dynamically
concentrated in this structure; pointings A, X, B, and H have O
(Ne) abundances that are 3-8 (2-5) times higher than ISM values
(Figure 6, right). For other regions, both the temperature and the
abundances are more similar to those for the SNR shell emission
component, indicating that any ejecta emission is dominated by the
projected emission from the shell.

Given that the SNR shell shows both variations in brightness and
enhanced abundances of Ne, it is not impossible that the emission
we are associating with an ejecta component actually corresponds
to anomalous thermal regions from the shell, seen in projection.
However, the good correspondence between this thermal component and
the nonthermal emission, which is not seen anywhere else along the
SNR shell, makes this extremely unlikely. The enhanced abundance
of Mg seen in individual fits to Region 1 in Pointing A support
this picture as well.

The variations in both PWN emission components are illustrated in
Figure 7 (right) where we compare MOS2 spectra from three distinct
regions within Vela X (indicated by dashed circles in 
Figure 6). It is clear, for example, that the
nonthermal contribution in region H1 is much lower than in A1,
despite the thermal flux being similar, albeit with a lower temperature
in H1.  Region G5 shows a significantly lower temperature. Also
evident is a significant variation in the strength of the emission
lines. Variations in the ratio of O~VIII to O~VII are easily observed.

The relative strengths of the O VII and O VIII lines within Vela~X
are illustrated in the line-to-continuum maps shown in Figure 8.
The maps were created by extracting images in the energy ranges
$480 - 615$~keV and $615 - 715$~keV for the two respective lines,
and estimating the continuum in each line region through an
interpolation of the continuum regions in the range $425 - 475$~keV
and $1120 - 1160$~keV range, assuming a power-law behavior for the
continuum over this range. The regions with enhanced O~VIII emission
relative to that from O~VII are seen to correlate with the temperature
map in Figure 6 (center). This is expected, with the lower temperature
plasma producing weaker emission from the higher ionization state, but
is also consistent with the cooler SNR component dominating the
ejecta emission in these regions.

\section{Modeling}

Both the morphology of the Vela SNR and estimates of the density
in regions around the remnant suggest that the surrounding density
is higher in the NE than in the southwest (SW). This is consistent
with the notion that the structure of Vela~X, with the bulk of the
PWN located to the south of the pulsar, is the result of the SNR
RS propagating more rapidly from the NE \citep{Blondin_etal01}. To
better understand the structure of Vela~X, it is necessary to
consider models for the evolution of the entire composite system.

\subsection{Analytical Modeling}

We note that multiple studies indicate that the medium surrounding
the Vela SNR contains clumps or small clouds \citep[e.g.,][]{NS04,
Miceli_etal05}, complicating our treatment of the density profile.
A similarity solution for the evolution of an SNR in a cloudy medium
was developed by \citet[][WL91 hereafter]{WL91} by introducing two
additional variables to the Sedov solution -- the mass ratio of
cloud material to intercloud material in the SNR ($C$), and the
ratio of the cloud evaporation timescale to the age of the SNR.
The evolution of the SNR radius is similar to that found from a
Sedov solution, but for a density that is weighted toward larger
values than that of the intercloud medium, due to contributions
from additional material evaporated from the engulfed clouds.  While
MHD  simulations that incorporate realistic cloud size distributions
and a full treatment of the thermal conduction effects show
considerable deviations from the WL91 solutions \citep{Slavin_etal17},
the associated behavior of the expansion law is still a reasonable
approximation.  This is illustrated in Figure 9 in which we plot
the radius as a function of time for three simulations by
\citet{Slavin_etal17} using different filling factors of clouds
with a factor of 100 density contrast with the intercloud medium
(shown as solid curves), For each, we plot (as dashed curves) Sedov
solutions using fixed density values shown in the Figure. While a
full treatment of a composite SNR evolving into a cloudy ISM
is beyond the scope of this paper, Figure 9 illustrates that evolution
at a fixed density (larger than that of the intercloud material,
but much smaller than that for the clouds) provides an adequate
description.

\begin{figure}[t]
\centering
\setlength\fboxsep{0pt}
\setlength\fboxrule{0.0pt}
\fbox{
\includegraphics[width=3.4in]{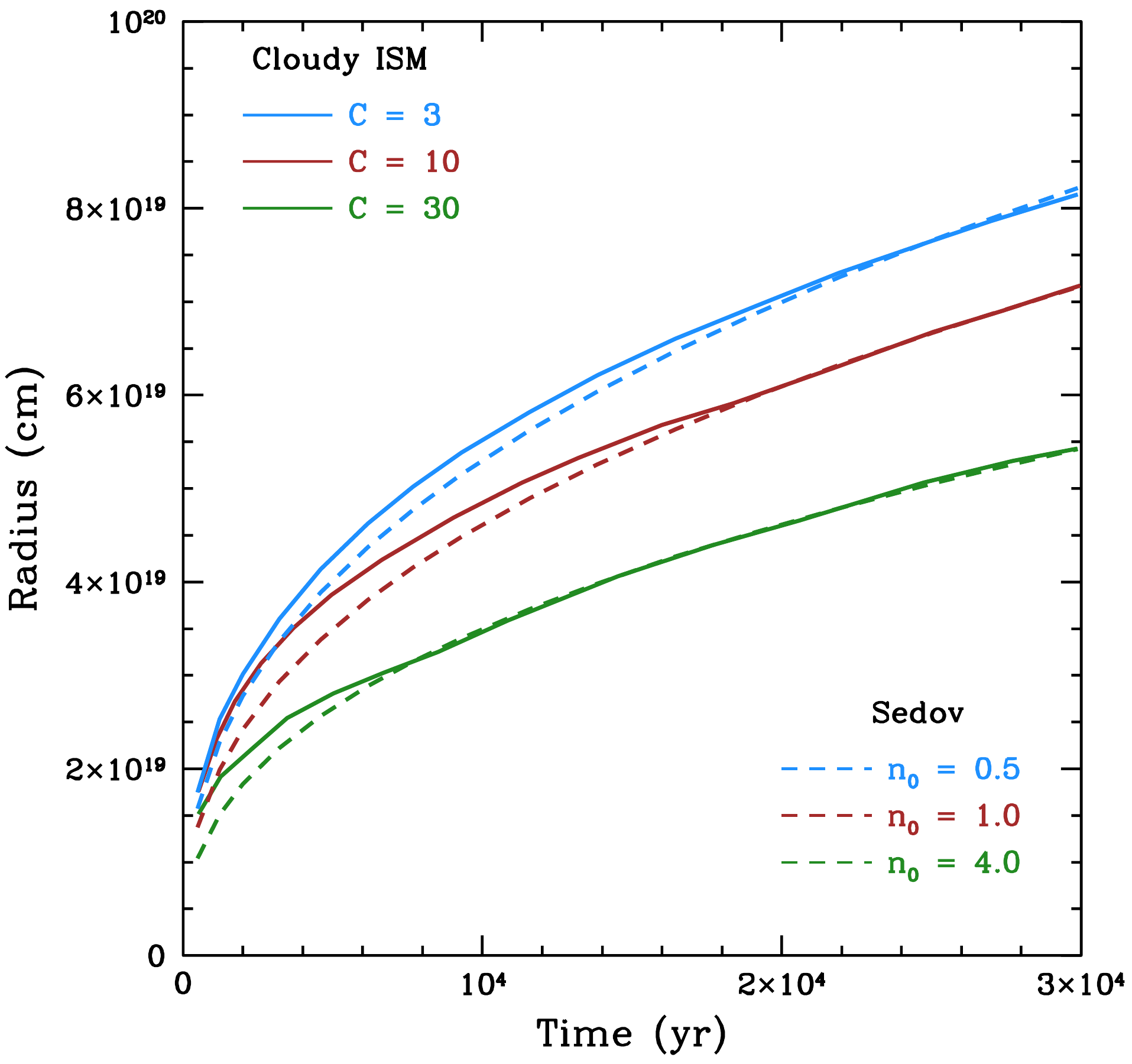}
}
\caption{
SNR radius evolution for expansion in a cloudy ISM (solid), and the
Sedov solution (dashed) for evolution into a uniform medium, both
for $E_{51} = 1$. The cloudy ISM curves (from Slavin et al. 2017)
assume density values of $0.25 (25) {\rm\ cm}^{-3}$ for the intercloud
(cloud) regions. The value of $C$ corresponds to the ratio of the
cloud mass to that of the intercloud medium. The Sedov solution
curves provide reasonable approximations assuming densities roughly
connected to the evaporated cloud density in the {\red cloudy ISM} model.
}
\label{fig9}
\end{figure}

\citet{Sushch_etal11} have simulated the evolution of the Vela SNR
using exactly this approach with the WL91 solutions. They assume a
higher concentration of clouds in the NE region of the SNR, leading
to a larger effective density in this direction than in the SW,
which they suggest may be due to the presence of a stellar wind
bubble blown by a nearby Wolf-Rayet star, in the $\gamma^2$ Velorum
binary system, creating a step-like density increase from the SW
to the NE. Models of the radio emission predicted from such a
scenario are able to reproduce the major observed radio properties
of the Vela SNR \citep{SH14}.  Here we make a similar assumption
about an overall density profile that increases from the SW to the
NE. 

An important constraint on the properties of this surrounding medium
is set by the morphology of Vela~X, where the RS appears to have
overtaken the pulsar from the NE direction. To investigate the basic
evolutionary parameters of the SNR, we have used solutions of
\citet{TM99} for the radius of the FS and the RS during evolution
from the ejecta-dominated phase to the Sedov phase.  While this
ignores the effects of the PWN, it provides a framework in which
to investigate the basic evolution for different values of the
ambient density, explosion energy, ejecta mass, and ejecta density
profile.

\begin{figure}[t]
\centering
\setlength\fboxsep{0pt}
\setlength\fboxrule{0.0pt}
\fbox{
\includegraphics[width=3.4in]{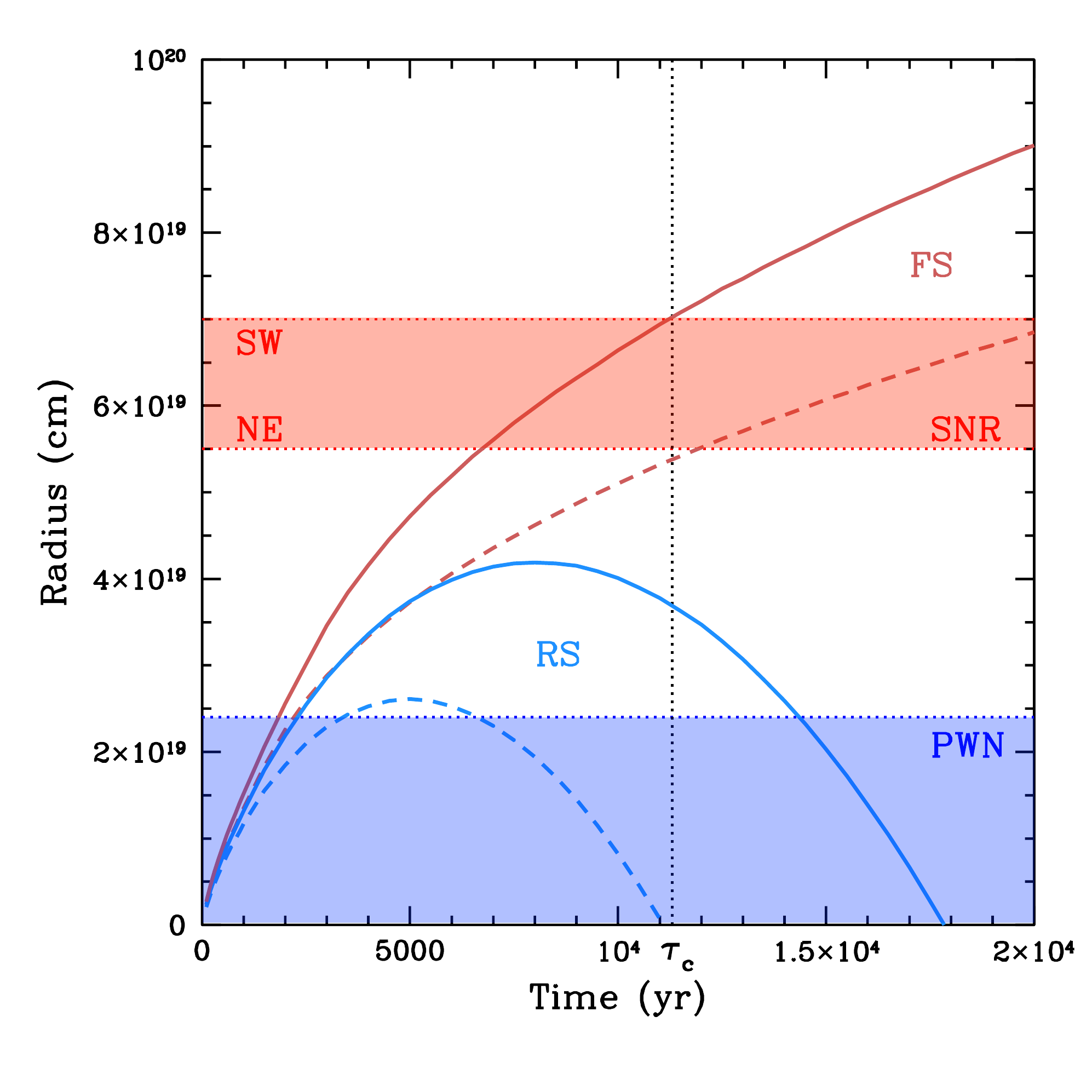}
}
\caption{
Forward (red) and reverse (blue) shock evolution with time using solutions
of Truelove \& McKee (1999). The solid (dashed) curves correspond to ambient
densities of $n_0 = 0.15 (0.5) {\rm\ cm}^{-3}$ with $E_{51} = 1$, $M_{ej}
= 8 M_\odot$, and a power law index $n = 12$ for the outer ejecta.
}
\label{fig10}
\end{figure}

In Figure 10 we plot solutions for the FS (red) and RS (blue) radius
for two different ambient densities. We have chosen values that
yield the observed FS radius in the NE and SW at $t = \tau_c$, the
characteristic age of the pulsar. We have assumed an ejecta profile
with a constant-density core that transitions to a steep power law
at larger radii \citet{Chevalier82}.  We have used a power law index
$n = 12$, which is typical of core-collapse supernovae
\citep[e.g.,][]{Chevalier82, MM99}, although the profile is quite
similar for $n=9$.  We have adjusted the ejecta mass until we
obtained solutions that provide a RS radius that has evolved
completely back to the SNR center from the NE direction. The RS
from the SW direction would, in this case, not yet have reached the
southern portions of the PWN (whose extent is indicated by the
dotted blue line), a result that does not hold with more detailed
modeling (see below).

We note that, for different values of the braking index and the
initial spin-down power, the actual age of the Vela Pulsar (and its
SNR) may be considerably larger than $\tau_c$. With no definitive
value for the system age, here and below we use $\tau_c$ as a
convenient age estimate to investigate the structure of Vela~X.
Similar results can be obtained using different values for these
input parameters, accompanied by modifications of the ambient
density. 

\begin{figure*}[t]
\centering
\setlength\fboxsep{0pt}
\setlength\fboxrule{0.0pt}
\fbox{
\includegraphics[width=\textwidth]{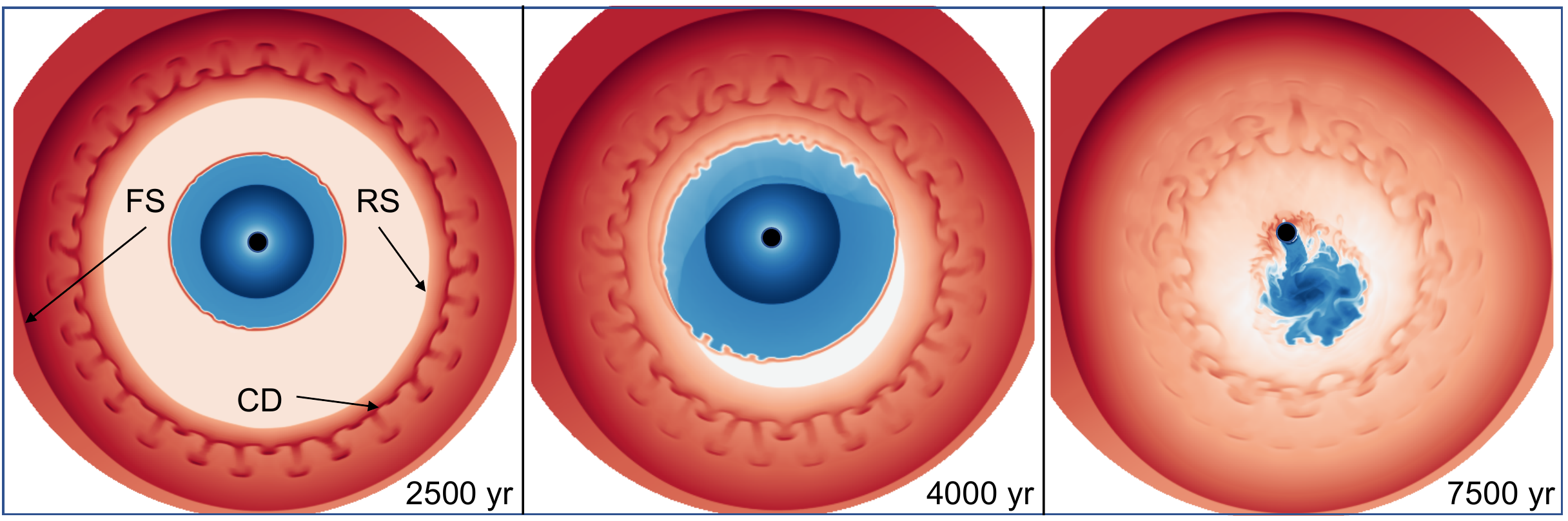}
}
\caption{
Hydrodynamical simulations of a composite SNR evolving into a density
gradient. Each panel shows the density at a different stage of the
evolution, with the lowest density regions (blue) corresponding to
regions dominated by relativistic gas from the pulsar. The FS, RS,
and contact discontinuity (CD) for the SNR are indicated in the 
left panel. The PWN
becomes disrupted as the SNR RS propagates more rapidly from the
NE, due to the external density gradient, eventually forming a relic
nebula connected by a trail of emission from the pulsar. See text
for further discussion.
}
\label{fig11}
\end{figure*}

\subsection{Hydrodynamical Modeling}

In an effort to understand the overall structure of Vela~X in the
context of the basic evolutionary picture described above, with
particular interest in the cocoon region of the PWN, we have carried
out modeling of the system using simulations with the VH-1 hydrodynamics
code as modified to treat the evolution of an SNR expanding into
an inhomogeneous medium, and with a central pulsar creating a PWN
that expands into the interior ejecta. The model is based on that
from \citet{Blondin_etal01}, but also includes treatments for following
the age of the particles injected by a moving pulsar and for tracking
the mixing of ejecta into both the swept-up CSM and the 
PWN material \citep{Kolb_etal17}.

We modeled the pulsar wind as a high Lorentz factor wind from the
pulsar, which we subsequently refer to as a relativistic component,
though it is treated here as a $\gamma = 4/3$ fluid injected at the
pulsar position (initially at the SNR center) with $\dot{M} =
2\dot{E}/v_w^2$. For computational purposes, the wind speed $v_w$
was taken as $\sim 1.5 \times 10^{9}{\rm\ cm\ s}^{-1}$, considerably
lower than the actual electron injection speed for a PWN. This results in a
high effective mass of electrons, but has no significant impact on
the subsequent evolution (Kolb et al. 2017). The pulsar position
was initially taken to be at the center of the SNR. In our initial
2D simulations, the pulsar was given a velocity of $61 {\rm\ km\
s}^{-1}$ in a direction consistent with the observed proper
motion.\footnote{In the simulation, the pulsar is kept stationary
at the center of the grid, while the rest of the grid is given a
velocity opposite that of the pulsar proper motion.} 

The pulsar
spin-down power is evolved as
\begin{equation} 
\dot{E}(t) = \dot{E_0}(t) \left(1 + \frac{t}{\tau_{sd}}\right)^{-\frac{n+1}{n-1}} 
\end{equation} 
where
$\tau_{sd}$ is the spin-down timescale and $n$ is the braking index.
While measurements of the braking index for many middle-aged pulsars
like the Vela pulsar are complicated by the presence of frequent
glitches, \citet{Lyne_etal_1996} and \citet{Espinoza_etal17} have 
measured $n = 1.7 \pm 0.2$,
well below the value $n = 3$ expected for standard magnetic dipole
spin down.  Younger pulsars show braking indices closer to the
dipole value, suggesting the possibility that extensive glitching
behavior develops later, somehow connected with a change in the
index. Increases in the superconducting component of the NS
moment of inertia can lead to $n < 3$, as can a temporal increase
in the surface magnetic field (as might occur through diffusion of
a field initially buried under a non-magnetic crust formed from early
fallback onto the NS) \citep[e.g.,][]{Page13}.  For our simulations,
we have investigated both $n = 1.7$ and $n = 3$ and find that, in
general, a smaller value for $n$ can be compensated by an increase 
in the value of $\tau_{sd}$, which would also result in a spin
period at the current epoch that does not differ much from the period
at birth.

We have treated the inferred density
gradient into which the Vela SNR appears to be expanding by 
\begin{equation} 
\rho(z) = \rho_0\left[1 - \frac{2-x}{x} \tanh(z/H)\right].  
\end{equation} 
This represents a step-like
transition between two constant-density regions, where $z$ is the
coordinate in the direction of the gradient, $H$ is the size scale
for the density transition, and $x$ regulates the amplitude of the
density step.  We choose parameters that yield a density contrast
of four across the SNR, as suggested by measurements described
above. The input parameters for the HD model are summarized in Table
2.

\begin{table}[b]
\begin{center}
\caption{\footnotesize{HD Model Parameters}}
\label{tab:mod}
\begin{tabular}{llc}
\toprule
\noalign{\smallskip}
\noalign{\smallskip}

Parameter & Description & Value \\ \hline
\\
Input Parameters \\
$v_p ({\rm\ km\ s}^{-1})$ 	& Pulsar velocity 		& 61 \\
$v_p$ direction (deg)			& 				& $31$\\
$E_{51} (10^{51}$~erg) 		& Explosion energy 		& 1 \\
$M_{ej} (M_\odot)$ 		& Ejecta Mass 			& 5 \\
$n_{ej}$ 			& Ejecta density profile index 	& 12 \\
$\dot{E}_0 ({\rm\ erg\ s}^{-1})$ & Initial spin-down power 	& $10^{39}$ \\
$n$ 				& Braking index 		& 3.0 \\
$\tau_0$ (yr) 			& Spin-down timescale 		& 1000\\
$n_{0,min} ({\rm cm^{-3}})$ 	& Minimum ambient density & 	0.18  \\
$x$ 				& $n$ contrast parameter$^a$ & 1.25\\
$H$~(pc) 			& $n$ contrast length scale & 6.5 \\
$n_0$ gradient (deg)		& North of west 		& $121$ \\
\\
Simulation Output \\
$R_{\rm SNR,NE}$ (cm) 		& SNR Radius (NE)		& $5.7 \times 10^{19}$\\
$R_{\rm SNR,SW}$ (cm) 		& SNR Radius (SW)		& $7.1 \times 10^{19}$\\
$l_{\rm cocoon}$ (cm) 		& Cocoon length 		& $1.8 \times 10^{19}$\\
$\dot{E} ({\rm\ erg\ s}^{-1}$) 	& Current spin-down power 	& $6.6 \times 10^{36}$\\

\noalign{\smallskip}
\hline
\end{tabular}
\end{center}
$a$ Parameter required to produce density contrast of four from SW to NE.
\end{table}

Our initial simulations, in which the parameter space was explored,
were carried out in 2D.  In Figure 11 we show density profiles of
the composite system at three discrete time steps in order to
illustrate the overall evolution. The simulation uses a cylindrical
2D grid with 600 zones in both the $r$ and $\phi$ directions. The
circle in the center represents the position of the pulsar, and the
surrounding light-blue structure is the PWN. The external density
increases from SW to NE as described by Equation 2. The dark inner
region represents the unshocked wind, and the termination shock is
at the boundary of these two regions. The FS is shown in red and
the RS is orange in color. The ISM and ejecta regions are separated
at the contact discontinuity where Rayleigh-Taylor instabilities
are evident.

At an age of 2500 years, the asymmetric propagation of the RS is
evident, with the shock approaching the PWN from the NE, but still
far from the nebula in the SW. By 4000 years the RS has encountered
the PWN in the NE, sweeping PWN gas back toward the location
of the pulsar. By an age of 7500 years the entire PWN has been
disrupted by the RS, with the component arriving from the NE
completely sweeping over the pulsar, creating an apparent trail of
emission to the south.

In Figure 12 we show the system at an age of 11.3~kyr, matching the
Vela Pulsar characteristic age, here in a 3D simulation using the
parameters arrived out for our 2D modeling. The 3D simulation was
computed on a spherical Yin-Yang grid \citep{whm10} using 384 radial
zones and an angular resolution of 56.25$^\prime$. The pulsar has
been given a kick of $80.2 {\rm\ km\ s}^{-1}$  such that the
velocity component along the plane in Figure 12 is 61 km/s and the
component along line of sight is $52 {\rm\ km\ s}^{-1}$ away from
the observer.  The projected direction of the pulsar motion as well
as the density gradient are indicated with arrows. At this stage
the RS has swept the pulsar wind far to the SW of the pulsar,
creating a relic PWN infused with both relativistic gas and ejecta.
The density scale (in ${\rm g\ cm}^{-3}$) is shown in the legend,
and the outer radio contour for Vela X (scaled in size assuming a
distance of 290~pc) is shown for comparison.  Red/orange regions
are primarily ISM or ejecta while the low-density blue regions have
large concentrations of relativistic gas.  The dashed box region
is shown in an extended panel on the right.  The black contour marks
the boundary of the PWN, composed of both relativistic gas and
mixed-in ejecta; material outside the contour is pure ejecta (or
ISM in the outermost regions). Within the PWN region, turbulent
vortex-like structures have created a stream of gas between the
pulsar and the relic PWN whose properties appear very similar to
the Vela X cocoon (discussed further below). The elongated structure
is characterized by a dense filament of ejecta aligned with a similar
structure of relativistic gas. The length scale of the structure
is $\sim 1.5 \times 10^{19}$~cm, corresponding to an angular size
of $\sim 57$~arcmin at the distance of Vela~X -- slightly larger,
but in overall agreement with the $\sim 45$~arcmin length of the
Vela~X cocoon.

\section{Discussion}

The gas distribution in the region of the disrupted PWN is a complex
mix of relativistic particles from the pulsar and ejecta material
swept up by the SNR RS. Figure 13(a) reproduces the total gas density
in this region, with the legend indicating the scale in ${\rm g\
cm}^{-3}$. As in Figure 12, low-density blue regions are dominated
by relativistic material injected by the pulsar. Chaotic,
elongated structures of ejecta and relativistic gas comprise the
cocoon structure, while the relic PWN persists in the southern-most
regions.  Large amounts of ejecta are found throughout most of the
structure, consistent with our detection of thermal gas with enhanced
abundances in the spectra from Vela~X. Regions along the cocoon
have high density, presumably producing the bright elongated
structures seen in Figure 3. These structures are similar to those
first discussed by \cite{cr11}. Interestingly, while the innermost
ejecta are expected to have a significant Fe component, the abundances
derived from our spectral investigations are consistent with those
expected to be found farther from the ejecta core. This may be
indicative of significant mixing of the inner ejecta early in
the evolution of the supernova that formed the system.

Figure 13(b) shows the density of
just the relativistic gas in the hydro simulation, making it clear
that one might expect synchrotron radiation from the entire PWN
structure.  The density along the cocoon-like structure is higher
than in the immediate surroundings, but not significantly higher
than in some other parts of the PWN.

\begin{figure}[t]
\centering
\setlength\fboxsep{0pt}
\setlength\fboxrule{0.0pt}
\fbox{
\includegraphics[width=3.3in]{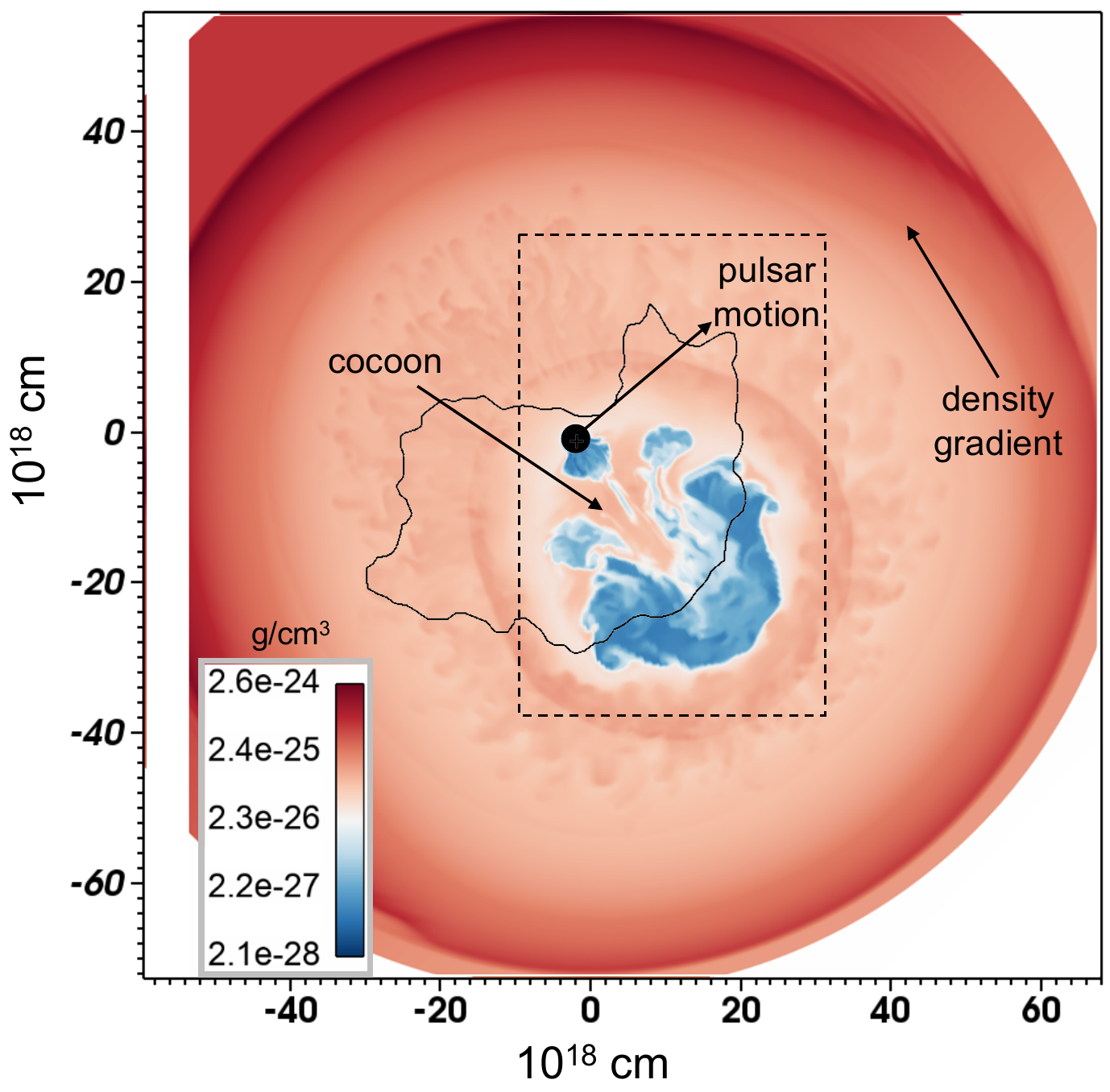}
}
\caption{
Simulation at the characteristic age of the Vela pulsar.
The plot on the left represents the density, with the
lowest density regions (blue) corresponding to regions dominated
by relativistic gas from the pulsar. Note the channel of thermal gas
beyond the wind immediately below the pulsar, extending into the
relic PWN, corresponding to a structure similar to the Vela~X cocoon.
The black contour represents Vela~X at a distance of 290~pc. The
region within the dashed box is shown in Figure 13.
}
\label{fig12}
\end{figure}

\begin{figure*}[t]
\centering
\setlength\fboxsep{0pt}
\setlength\fboxrule{0.0pt}
\fbox{
\includegraphics[width=\textwidth]{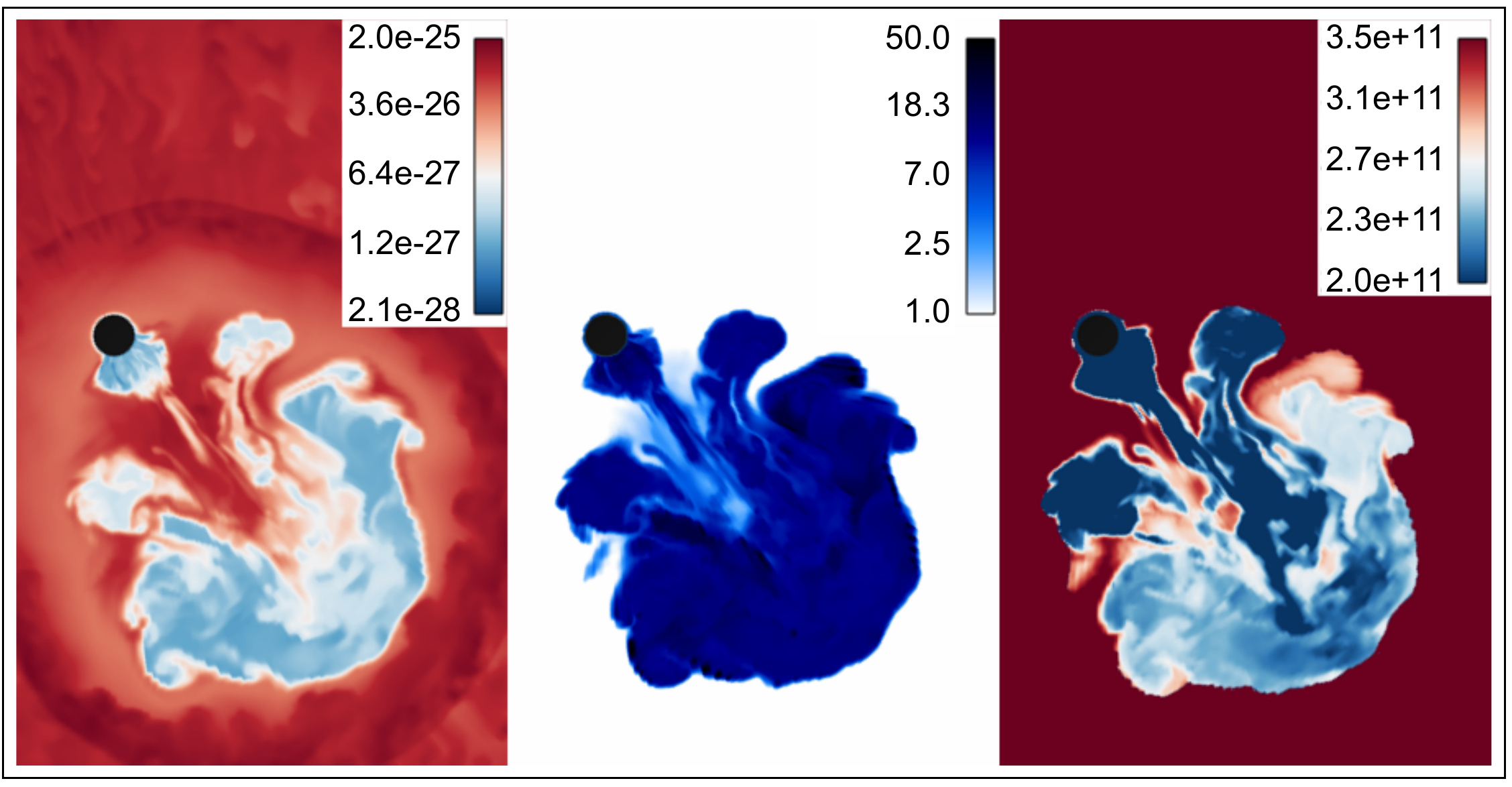}
}
\caption{
Inner regions of the Vela~X HD simulation. Panel (a) shows the
density (in ${\rm\ g\ cm}^{-3}$). Here, red/orange regions have
higher density and are dominated by ejecta, while blue regions are
dominated by low-density PWN gas. Panel (b) shows the relative
effective density of just the relativistic (PWN) gas.  Panel
(c) plots the age of the relativistic gas (in seconds since the gas
entered the grid). Note that the scale is saturated at small 
timescales to emphasize the distribution of younger particles along
the cocoon.
}
\label{fig13}
\end{figure*}

While we are unable to calculate the magnetic field strength
within the evolved PWN from our purely hydrodynamical simulations,
it is instructive to investigate the age of the relativistic
gas in different regions of the nebula in order to predict
the general behavior of the associated synchrotron radiation.
The characteristic synchrotron energy of an electron is given by
\begin{equation}
h\nu_{\rm s} \approx 2.2 E_{e,100}^2 B_{10}{\rm\ keV},
\end{equation}
where $E_{e,100}$ is the electron energy in units of 100~TeV and
$B_{10}$ is the magnetic field strength in units of 10$\mu$G, and
the lifetime against synchrotron losses is
\begin{equation}
\tau_{syn} \approx 820 E_{e,100}^{-1} B_{10}^{-2}{\rm\ yr}.
\end{equation}
For emission at 1~keV, and assuming a magnetic field strength of 5$\mu$G
based estimates for Vela~X (LaMassa et al. 2008), the synchrotron lifetime
of the electrons emitting X-rays in the soft band is $\sim 3.5$~kyr
($\sim 1.1 \times 10^{11}$s). The age (in seconds) of the relativistic gas in
the PWN (i.e., the time since injection into the nebula) is shown
in Figure 13(c). Particles in the saturated (dark blue) regions nearest the
pulsar have ages less than $\tau_{syn}$, while those in the southern reaches
of the cocoon and in the relic nebula have ages longer than the 
synchrotron loss time. This appears consistent with the observed
variation in spectral index in Vela~X, where the spectrum is hard
in the inner regions but softer at large distance from the pulsar
(see Figure 7).

The X-ray emission from pointing I, which overlaps the position at
which the peak GeV emission is seen in \fermi~LAT observations,
shows a harder nonthermal component than regions at similar distances
on the opposite side of the cocoon. This is difficult to reconcile
with the observed \gamray\ properties of Vela~X which, along with
the radio and X-ray data, seem to indicate two distinct electron
populations, with the GeV emission being associated with the
low-energy component \citep{deJager_etal08}. \cite{hinton_etal11}
suggest that diffusion of particles from the relic nebula, which
is proposed to contain an older population of particles, can explain
both the observed steep GeV spectrum of Vela~X and the observation
that the GeV emission is more broadly distributed than the TeV
emission. The harder X-ray spectrum at the GeV peak seems to contrast
with this suggestion, although the considerable hydrodynamical
transport of the PWN gas is observed to form pockets of younger gas
within regions of older particles (Figure 13c). MHD simulations
that can treat the particle diffusion along with the overall evolution
of the magnetic field within the PWN are required to address this
issue further. We note, also, that the jet from the Vela Pulsar shows
a complex structure that extends toward the general direction of
this region \citep[e.g.,]{Pavlov_etal_2003}, although the scale of
this structure appears quite small relative to the distance to
this region.

Finally, we note that Vela~X is considerably broader in directions
perpendicular to the cocoon than the extent in our simulations,
which we attribute to a more complex ambient density structure than
the assumed smooth jump in a single direction. In addition, while
the RS sweeps the entire relic PWN to the south of the pulsar in
our simulations, we note that emission from Vela~X is observed
somewhat to the north of the Vela Pulsar. This may be a result of
magnetic effects that are not properly modeled in the purely
hydrodynamic simulations.

\section{Conclusions}

We have carried out X-ray observations and hydrodynamical simulations
of Vela~X in an effort to understand its overall structure in the
context of the evolution of a PWN within the confines of an SNR.
Our results show that the properties of Vela~X are well described
by a model in which the PWN has undergone asymmetric disruption
from the SNR RS, resulting from the Vela SNR evolving in a non-uniform
ISM with a higher density in the NE, as originally suggested by
Blondin et al. (2001). Our large-scale mapping of key regions within
Vela~X provides important details of the overall structure that
confirm earlier measurements of both thermal and nonthermal emission
\citep{MO97,LaMassa_etal08}. We produce temperature, abundance, 
spectral index, and equivalent width  maps that reveal significant
variations in the properties of the thermal and nonthermal gas within
Vela~X. We find the following:

1. At an age equal to the characteristic age of the pulsar, the
overall size of the SNR and PWN can be approximately reproduced
assuming evolution into a density gradient whose values are constrained
by previous density measurements for the Vela SNR.

2. The interior of Vela~X contains both relativistic gas, producing
synchrotron radiation, and ejecta material whose thermal emission 
establishes enhanced abundances of Ne, O, and Mg. This is consistent
with HD simulations that show significant mixing of ejecta into the PWN 
during the RS-crushing phase of evolution.

3. Equivalent-width maps show a complex variation in the relative
contributions of O~VII and O~VIII to the emission. Both broad fitting
of the ejecta temperature and modeling of spectra from discrete
regions show that these variations are associated with temperature
differences that are primarily associated with regions where the
SNR shell component dominates emission from the ejecta.

4. The Vela~X ``cocoon'' consists of an extended region whose
emission is from a combination of ejecta and PWN material.
High-resolution X-ray images provide some indication of fine-scale
filamentation, but show that the structure is largely diffuse on
small scales. Simulations show that such a structure can naturally
be formed as part of the hydrodynamical development of the RS-PWN
interaction. 

5. The spectrum of the nonthermal emission is harder near the pulsar
than at regions farther along the cocoon, and beyond. The HD
simulations indicate that this is consistent with both the cocoon
and the relic PWN being dominated by particles injected by the
pulsar at an earlier age, now subject to synchrotron losses, while
the region nearer the pulsar contains particles more recently
injected by the pulsar.

6. The nonthermal X-ray spectrum at the peak position of the
GeV emission observed with the \fermi-LAT appears harder than
emission at similar offset distances from the pulsar and the
cocoon.

The overall picture from our observations and simulations are
consistent with the picture that the overall morphology of 
Vela~X is the result of an asymmetric RS interaction associated
with expansion of the SNR into medium whose density is higher
in the NE regions. The cocoon appears to result from the RS 
approaching and displacing the PWN from the NE, then wrapping
around the highest pressure regions of the nebula close to the
pulsar, finally coming together on the opposite side of the
pulsar and over-pressuring the entrained plasma there. 

There remain important considerations for which extension of our
modeling to MHD will be necessary. In particular, since our HD
simulations do not treat the evolution of the PWN magnetic field,
our knowledge of the synchrotron losses suffered by particles in
different regions of the PWN is quite incomplete.  This, along with
magnetic effects that might dominate the dynamics on some spatial
scales, will be important to consider in future works in order to
address the overall distribution of the emission observed in the
GeV and TeV bands.

\acknowledgments
P.S. acknowledges partial support from NASA grants NNX09AP99G, 
GO1-12100X, and TM6-17002X, and NASA Contract NAS8-03060. In addition, 
he gives thanks to the Aspen Physics center, under whose hospitality portions
of this work were completed.  The authors would like to thank Douglas
Bock for providing the radio image shown in Figure 2, and Iurii Sushch
for helpful discussions about the Vela SNR system. We also thank the
anonymous referee for a careful review of the manuscript and for helpful
suggestions that have improved the presentation of the work.




\bibliographystyle{aa} 
\bibliography{bib_pos}

\end{document}